\documentclass{article}

\usepackage{complex-systems}

\pdfoutput=1

\usepackage{epsfig}
\usepackage{graphics}
\usepackage{graphicx}
\usepackage[centertags]{amsmath}
\usepackage{amsfonts}
\usepackage{amssymb}
\usepackage{url}
\usepackage{subfigure}
\usepackage{hyperref}

\begin{document}

\title{On soliton collisions between localizations in complex ECA: Rules 54 and 110 and beyond}

\author{
\authname{Genaro J. Mart{\'i}nez}
\\ [2pt]
\authadd{Departamento de Ciencias e Ingenier{\'i}a de la Computaci{\'o}n,}\\
\authadd{Escuela Superior de C\'omputo, Instituto Polit\'ecnico Nacional, M\'exico}\\
\authadd{Unconventional Computing Center, Computer Science Department,}\\ 
\authadd{University of the West of England, Bristol BS16 1QY, United Kingdom}\\
\authadd{\url{genaro.martinez@uwe.ac.uk}}%
\\ [2pt]
\and
\authname{Andrew Adamatzky}
\\[2pt] 
\authadd{Unconventional Computing Center, Computer Science Department,}\\ 
\authadd{University of the West of England, Bristol BS16 1QY, United Kingdom}\\ 
\authadd{\url{andrew.adamatzky@uwe.ac.uk}}%
\\[2pt]
\and
\authname{Fangyue Chen}
\\[2pt] 
\authadd{School of Sciences, Hangzhou Dianzi University}\\ 
\authadd{Hangzhou, Zhejiang 310018, P. R. China}\\ 
\authadd{\url{fychen@hdu.edu.cn}}%
\\[2pt]
\and
\authname{Leon Chua}
\\ [2pt] 
\authadd{Electrical Engineering and Computer Sciences Department} \\ \authadd{University of California at Berkeley, California, United States of America}\\ 
\authadd{\url{chua@eecs.berkeley.edu}}%
}


\maketitle

\markboth{Genaro J. Mart{\'i}nez et al.}{On soliton collisions in complex ECA: Rules 54 and 110 and beyond}

\begin{abstract}
In this paper we present a single-soliton two-component cellular automata (CA) model of waves as mobile self-localizations, also known as: particles, waves, or gliders; and its version with memory. The model is based on coding sets of strings where each chain represents a unique mobile self-localization. We will discuss briefly the original soliton models in CA proposed with {\it filter automata}, followed by solutions in elementary CA (ECA) domain with the famous universal ECA {\it Rule 110}, and reporting a number of new solitonic collisions in ECA {\it Rule 54}. A mobile self-localization in this study is equivalent a single soliton because the collisions of these mobile self-localizations studied in this paper satisfies the property of solitonic collisions. We also present a specific ECA with memory (ECAM), the ECAM Rule $\phi_{R9maj:4}$, that displays single-soliton solutions from any initial codification (including random initial conditions) for a kind of mobile self-localization because such automaton is able to adjust any initial condition to soliton structures. \\

\end{abstract}

\section{Introduction}
\label{intro}
A {\it soliton} can be defined in an informal as follows: when two solitary waves travel in opposite directions and collide, they emerge after collision with the same shape and velocity asymptotically. The phenomenon of solitary wave first recognized by English engineer John Scott Russell \cite{kn:Russ44},\footnote{You can find original scanned Russell's papers from Eilbeck's website ``John Scott Rusell and the solitary wave'', \url{http://www.ma.hw.ac.uk/~chris/scott_russell.html}, 1998.} and first formalized by Diederick J. Korteweg and Gustav de Vries in 1895 \cite{kn:KV95}. However, in 1965 the physician Martin Kruskal coined the phenomenon of solitary wave as ``soliton.''

Solitons in one-dimensional (1D) CA have their own interest and history, they have been extensively studied since 1986 by Kennet Steiglitz and colleagues as you can see in \cite{kn:PST86, kn:SKW88, kn:JSS01, kn:RSP05}. This has been based on a variant of classic CA, known as {\it parity rule filter automata} (PRFA). A PRFA mainly uses newly computed site values as soon as they are available and they are analogous to Infinite Impulse Response (IIR) digital filters, while conventional CA correspond to Finite Impulse Response (FIR) \cite{kn:PST86}. Incidentally, yet more sophisticated solitons with PRFA were obtained by Siwak in \cite{kn:Siw02} showing large and multiple simultaneous solitonic collisions sequentially (i.e. not parallel mapping). We can also see, turbulence solitons in 1D CA explored by Aizawa, Nishikawa, and Kaneko in \cite{kn:ANK90}. 

Studies of 1D soliton CA are important because it allows for fast-prototyping of soliton logic. For practical implementations of soliton logic, see overview developed by Blair and Wagner in \cite{kn:BW02}, leads to novel designs of optical parallel computers. An interesting implementation showing the wave propagation equation in lattice gas simulated with a partitioned CA was developed by Margolus, Toffoli, and Vichniac in \cite{kn:MTV86, kn:TM87}. Solitons have found relevant and numerous applications, some of them are in: fiber optics, breather waves, non-linear Schr\"{o}dinger equation, magnets, and recently in proteins and DNA, bio-solitons \cite{kn:Dav90, kn:AA05, kn:MS08, kn:DP06, kn:HM03}.

Historically, complex CA have been related to the presence of mobile self-localizations (referred as well as: gliders, particles, or waves). The most famous CA is the two-dimensional (2D) CA Conway's {\it the Game of Life} \cite{kn:Gard70}, but we can also find a number of samples in 1D supporting mobile self-localizations, as we can see in \cite{kn:Ada94, kn:BNR91, kn:LPM07, kn:MMS06, kn:MNG04, kn:Piva07}. Some of them process explicitly signals (not mobile self-localizations) by Delorme and Mozayer in \cite{kn:DM02} or solve the firing squad synchronization problem by Umeo in \cite{kn:UK10}. Indeed, we can see how a number of CA have been exploited as physical models in \cite{kn:Marg84, kn:Wolf86, kn:Wolf86a, kn:CD98}.

This paper is organized as follows. Section~\ref{ca} gives a general introduction on CA and basic notation. Section~\ref{solitonsCA} presents experimental soliton solutions in CA including solitons in complex ECA rules 54 and 110, and we will reporting in this paper the soliton reactions emerging in Rule 54 from multiple collisions. Later in Sec.~\ref{Mem-solitons} we  displays a new ECAM able to solve experimentally the most simple single-soliton two-component solution from any initial configuration. Finally we will discuss some computing capacities based on solitons. In the last section we will discuss final remarks (Sec.~\ref{conclusions}).

\section{One-dimensional cellular automata}
\label{ca}

\subsection{Elementary cellular automata (ECA)}
\label{eca}

A CA is a quadruple $ \langle \Sigma,\varphi,\mu, c_0 \rangle$ evolving on a specific $d$-dimensional lattice, where each cell $x_i$, $i \in N$, takes a state from a finite alphabet $\Sigma$ such as $x \in \Sigma$. A sequence $s \in \Sigma^n$ of $n$ cell-states represents a string or a global configuration $c$ on $\Sigma$. We write a set of finite configurations as $\Sigma^n$. Cells update their states via an evolution rule $\varphi: \Sigma^{\mu} \rightarrow \Sigma$, such that $\mu$ represents a cell neighbourhood that consists of a central cell and a number of neighbours connected locally. There are $|\Sigma|^{\mu}$ different neighbourhoods and if $k=|\Sigma|$ then we have $k^{k^n}$ different evolution rules.

\begin{figure}[th]
\centerline{\includegraphics[width=2.5in]{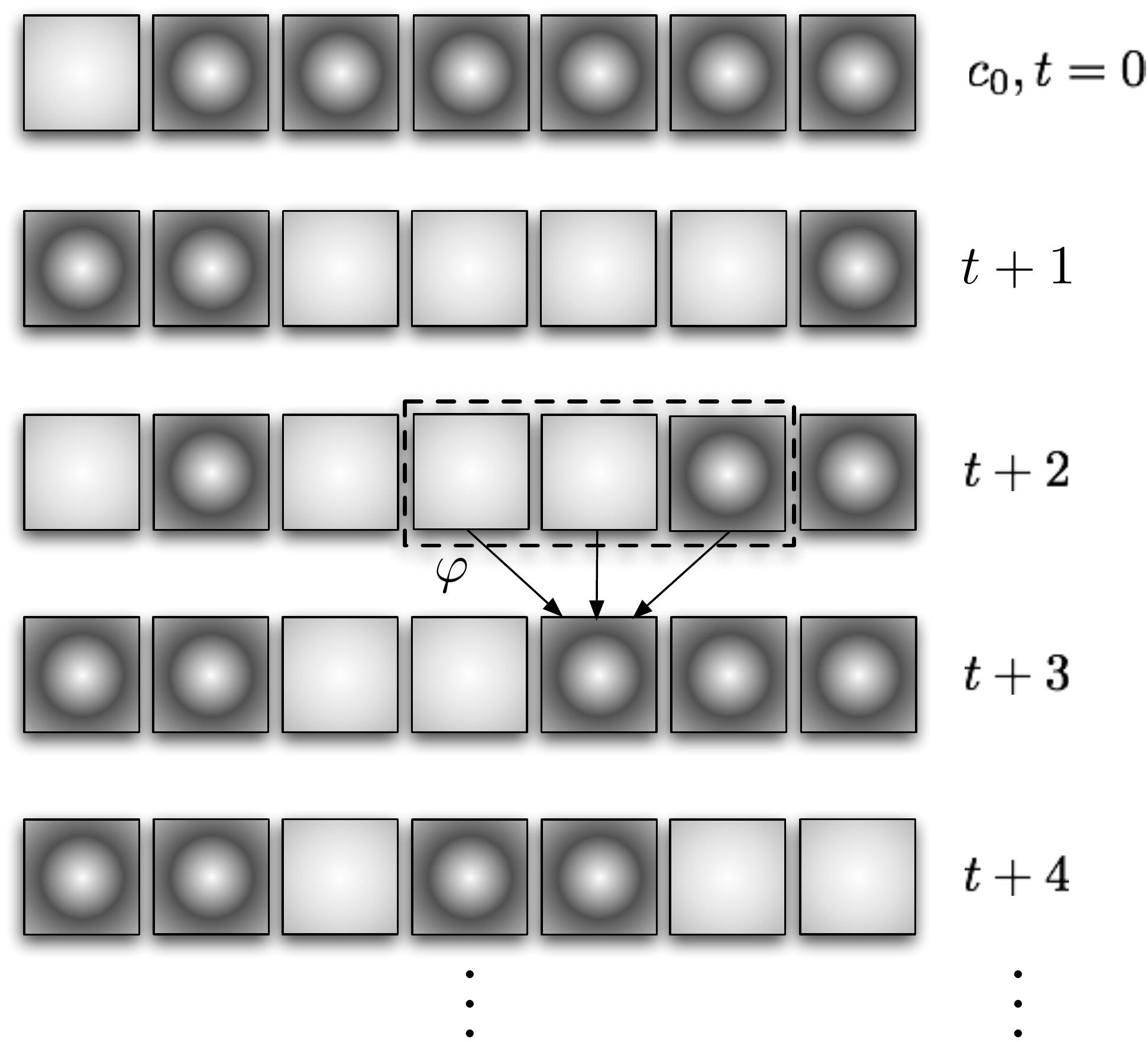}}
\caption{Dynamic in ECA on an arbitrary one-dimensional array and hypothetical evolution rule $\varphi$.}
\label{evolECA}
\end{figure}

An evolution diagram for a CA is represented by a sequence of configurations $\{c_i\}$ generated by the global mapping $\Phi:\Sigma^n \rightarrow \Sigma^n$, where a global relation is given as $\Phi(c^t) \rightarrow c^{t+1}$. Thus $c_0$ is the initial configuration. Cell states of a configuration $c^t$ are updated simultaneously by the evolution rule as:

\begin{equation}
\varphi(x_{i-r}^t, \ldots, x_{i}^t, \ldots, x_{i+r}^t) \rightarrow x_i^{t+1}.
\end{equation}

\noindent  where $i$ indicates cell position and $r$ is the radius of neighbourhood $\mu$. Thus, the elementary CA class represents a system of order $(k=2,\ r=1)$ (in Wolfram's notation \cite{kn:Wolf86a}), the well-known ECA. To represent a specific evolution rule we will write the evolution rule in a decimal notation, e.g. $\varphi_{R110}$. Thus Fig.~\ref{evolECA} illustrates how evolution dynamics work in one dimension for ECA.

\subsection{Elementary cellular automata with memory (ECAM)}
\label{ecam}

Conventional CA are memoryless:  the new state of a cell depends on the neighbourhood configuration solely at the preceding time step of $\varphi$. CA with memory are an extension of CA in such a way that every cell $x_i$ is allowed to remember its states during some fixed period of its evolution. CA with memory have been proposed originally by Alonso-Sanz in \cite{kn:AM03, kn:Alo06, kn:Alo09, kn:Alo11}.

Hence we implement a memory function $\phi$, as follows:

\begin{equation}
\phi (x^{t-\tau}_{i}, \ldots, x^{t-1}_{i}, x^{t}_{i}) \rightarrow s_{i} ,
\end{equation}

\noindent where $\tau < t$ determines the {\it degree of memory} and each cell $s_{i} \in \Sigma$ is a state function of the series of states of the cell $x_i$ with memory backward up to a specific value $\tau$. Later to execute the evolution we will apply the original rule on the cells $s$ as:

\begin{equation}
\varphi(\ldots, s_{i-1}, s_{i}, s_{i+1}, \ldots) \rightarrow x^{t+1}_i
\end{equation}

\noindent to get an evolution with memory. Thus in CA with memory,  while the mapping $\varphi$ remains unaltered, historic memory of all past iterations is retained by featuring each cell as a summary of its past states from $\phi$. We can say that cells canalise memory to the map $\varphi$~\cite{kn:Alo09}.

Let us consider the {\it memory function} $\phi$ as a {\it majority memory}, 

$$\phi_{maj} \rightarrow s_{i},$$

\noindent where in case of a tie given by $\Sigma_1 = \Sigma_0$ from $\phi$ hence we take the last value $x_i$. Thus, $\phi_{maj}$ function represents the classic majority function (for three values \cite{kn:Mins67}), then we have that:

\begin{equation}
\phi_{maj}(a,b,c) : (a \wedge b) \vee (b \wedge c) \vee (c \wedge a)
\label{majMem}
\end{equation}

\noindent that represents the cells $(x^{t-\tau}_{i}, \ldots, x^{t-1}_{i}, x^{t}_{i})$ and defines a temporal ring $s$ before getting the next global configuration $c$. Of course, this evaluation can be for any number of values of $\tau$. In this way, a number of functional memories may be used and not only majority, such as: minority, parity, alpha, $\ldots$, etc. (see \cite{kn:Alo09, kn:Alo11}).

\begin{figure}[th]
\centerline{\includegraphics[width=4.3in]{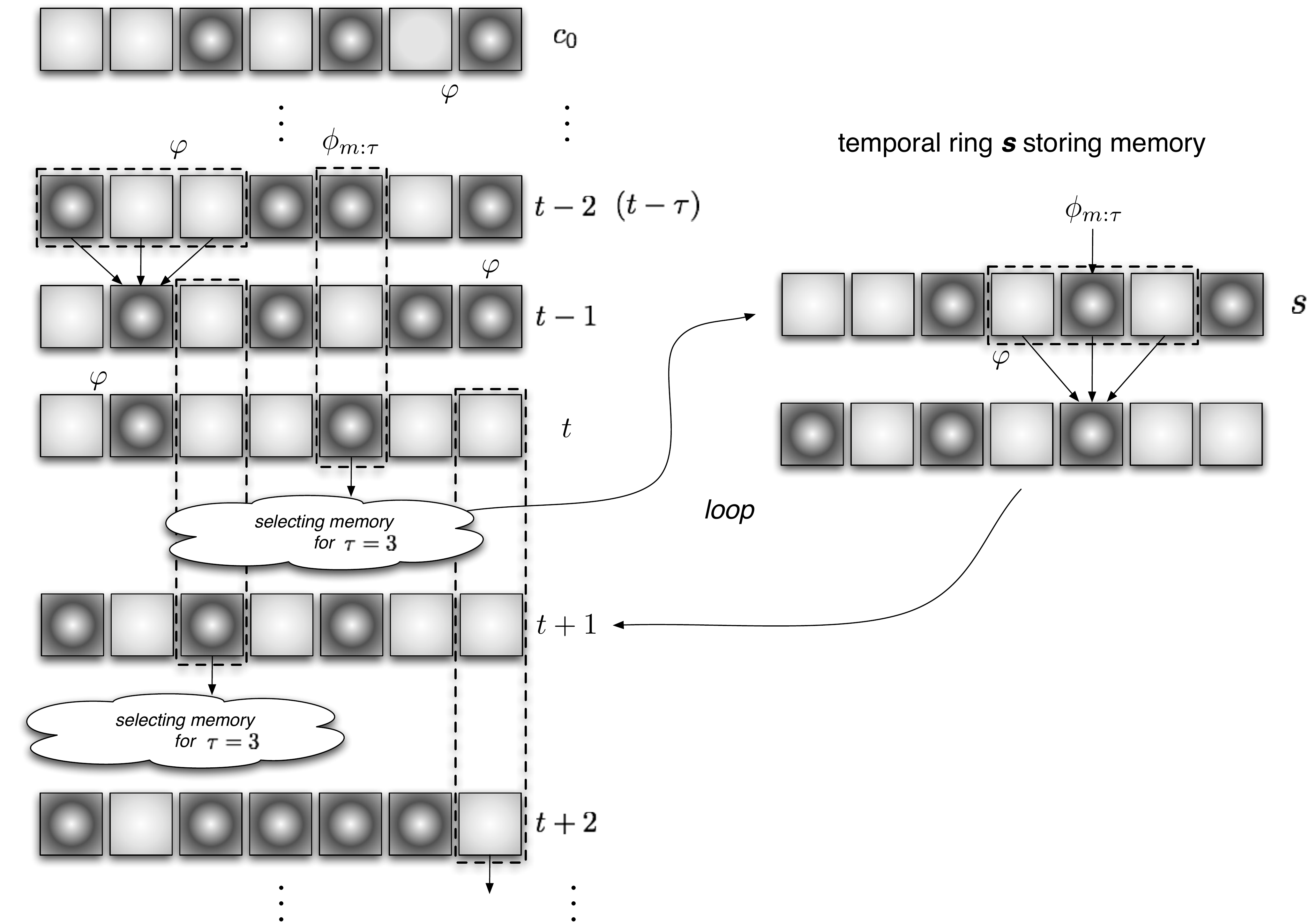}}
\caption{Dynamics in ECAM on an arbitrary one-dimensional array and hypothetical evolution rule $\varphi$ and memory function $\phi_m$ with $\tau=3$.}
\label{evolECAM}
\end{figure}

Evolution rules representation for ECAM is given in \cite{kn:MAA10, kn:MAS10, kn:MAS11}, as follows:

\begin{equation}
\phi_{CARm:\tau}
\end{equation}

\noindent where $CAR$ is the decimal notation of a particular ECA rule and $m$ is the kind of memory used with a specific value of $\tau$. This way, for example, the majority memory ($maj$) incorporated in ECA Rule 30 employing five steps of a cell's history ($\tau=5$)  is denoted simply as $\phi_{R30maj:5}$. The memory is functional as the CA itself, see schematic explanation in Fig.~\ref{evolECAM}. However, computationally a memory function has a quadratic complexity calculating its evolution space.

\section{Solitons in one-dimensional cellular automata}
\label{solitonsCA}

A {\it soliton} is a solitary wave with non-linear behaviour that preserves its form and speed, interacting with some kind of perturbation. The latter can be another wave or some obstacle, continuing its travel affecting only its phase and position since each collision. For example a water wave travelling and interacting with others waves, they can be found also in optics, sound, and molecules \cite{kn:Dav90}.

The solitary wave described by Scott become formally represented by the Korteweg-de Vries equation \cite{kn:KV95} as:

\begin{equation}
u_{t} + u_{xxxx} + uu_{x} = 0
\end{equation}

\noindent where the function $u$ measure high-wave and $x$-position at time $t$, and every subindex represent partial differences. Second term represent scattering-wave and the last term is the non-linear term \cite{kn:JSS01}.

However, we will indicate that soliton models related to CA do not find some direct relations matching some differential equation solutions. Nevertheless, Steiglitz has been displayed some properties with Manakov systems and PRFA in \cite{kn:Man73, kn:RSP05} in the search of computable systems collision-based soliton \cite{kn:SKW88}. In addition, Adamatzky in \cite{kn:Ada02a} has designed a way to manipulate solitons to implement logic gates. On the other hand, Chua has developed explicitly an extended analysis on how ECA can be described precisely as differential equations and cellular complex networks (CCN) in \cite{kn:Chua06}.

Although many studies were done in ECA here we cannot find much about the soliton phenomena for each rule. Complex ECA are direct candidates to explore such reactions from the interaction of their mobile self-localizations. Some explorations were described in \cite{kn:AM10, kn:ANK90, kn:BNR91, kn:CD98, kn:CSC, kn:LPM07, kn:MNG04, kn:Wolf86}. Solitons in CA are characterized as a set of cells self-organized emerging on the evolution space, such complex patterns have a form, volume, velocity, phase, period, mass, and shift. Of course, not all these mobile self-localizations may work as solitons because they depend on its interaction with other structures. Consequently a classification is necessary from the evolution space because they cannot be inferred from the local rule.

While a PRFA was designed to yield solitonic collisions calculating the new values as soon as they are available, their mobile self-localizations present a strong orientation to the left. This is a natural consequence of its function to calculate the next cell which evaluate the $(i-r)^{t+1}$ cells \cite{kn:PST86}. The main and most important difference with conventional ECA is that those mobile self-localizations working as solitons shall be searched explicitly and cannot be deduced to evolve the system. Thus not all complex ECA are able to produce collisions as solitons although they could evolve some kind of mobile self-localizations. Steiglitz has researched amply the PRFA with the goal of reaching unconventional computing devices based-soliton collision, as we can see in \cite{kn:SKW88, kn:Steig00, kn:JSS01, kn:RSP05}.

In the next subsections, we will discuss particular cases with complex ECA and ECAM, displaying exact codifications to get soliton collisions between mobile self-localizations and we will also present some computable capacities. In the ECA domain, we have selected and researched only complex rules 54 and 110, because none other ECA rules present a universe with such diversity of mobile self-localizations and consequently an ample diversity of collisions as well. In the ECAM domain we will present a single case that solve experimentally the most simple single-soliton two-component solution from any initial configuration, the evolution rule $\phi_{R9maj:4}$.

\subsection{Solitons in ECA Rule 110}
\label{110-solitons}

ECA Rule 110 is a complex cellular automaton evolving with a complicated system of mobile self-localizations. Its local function is defined as follows:

\begin{equation}
\varphi_{R110} = \left\{
	\begin{array}{lcl}
		1 & \mbox{if} & 001, 010, 011, 101, 011 \\
		0 & \mbox{if} & 000, 100, 111
	\end{array} \right. .
\end{equation}

\begin{figure}[th]
\centerline{\includegraphics[width=4.2in]{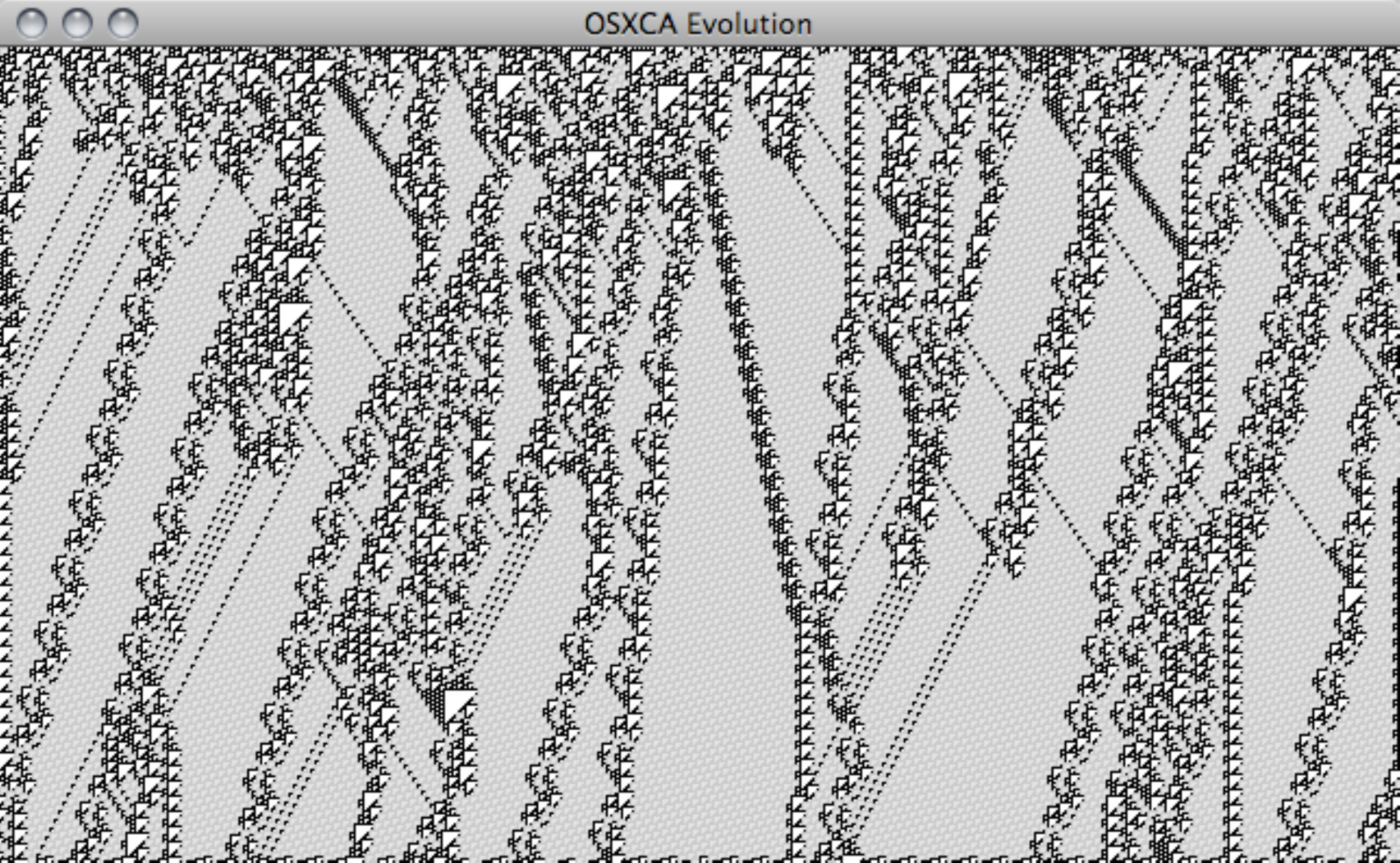}}
\caption{Random evolution in Rule 110 on a ring of 644 cells to 375 generations. White cells represent state 0 and black cells state 1 starting on a 50\% of density. A filter is selected to get a better view of mobile self-localizations on its periodic background.}
\label{evolR110}
\end{figure}

Figure~\ref{evolR110} illustrates the complex dynamics from a typical random initial condition selecting the evolution rule $\varphi_{R110}$. Here we can see how a number of mobile self-localizations emerge on its evolution space and how a number of them collide.

See detailed studies of Rule 110 and mobile self-localizations in: glider system\footnote{Gliders in Rule 110 \url{http://uncomp.uwe.ac.uk/genaro/rule110/glidersRule110.html}.} \cite{kn:MMS06, kn:MMS08, kn:Mc99}, universality \cite{kn:Cook04, kn:Wolf02, kn:Mc02, kn:NW06, kn:Cook08, kn:MMS11, kn:MAS11}, collisions and Rule 110 objects \cite{kn:MM01, kn:MMS07}. So generalities can be explored from the {\it Rule 110 repository}.\footnote{Rule 110 repository: \url{http://uncomp.uwe.ac.uk/genaro/Rule110.html}.}

\begin{figure}[th]
\centerline{\includegraphics[width=4.2in]{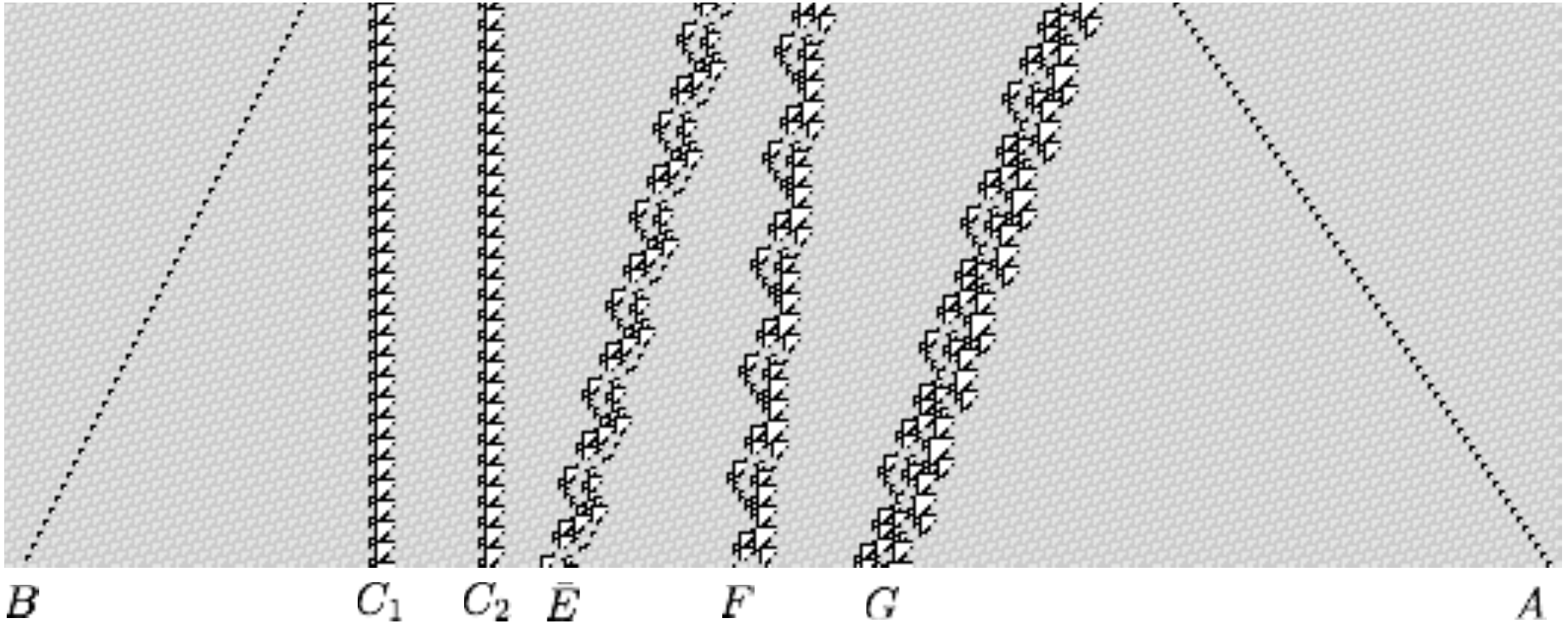}}
\caption{Set of mobile self-localizations with solitonic properties in Rule 110.}
\label{glidersSolitonR110}
\end{figure}

We will focus on mobile self-localizations that present solitonic reactions. Localizations that have such property are classified in Fig.~\ref{glidersSolitonR110}, following Cook's notation \cite{kn:Cook04}, here we can see stationary, shift-right, and shift-left (displacements) localizations.

Rule 110 has an unlimited number of collisions as a consequence of some extendible mobile self-localizations \cite{kn:MM01, kn:MMS06}. In this way, first we have constructed a set of configurations $c$ coding each localization and yielding the solitonic reaction desired.

To drive collisions and localizations, we will use the set of regular expressions `f$_1\_1$' localizations-based to code initial configurations in Rule 110, for full details please see \cite{kn:MMS08}.\footnote{Regular language glider-based in Rule 110: \url{http://uncomp.uwe.ac.uk/genaro/rule110/listPhasesR110.txt}.}

\begin{table}[th]
\centering
\small
\begin{tabular}{|c|c|c|c|c|}
\hline
mobile self-localization & shift & period & speed & volume \\
\hline \hline
$A$ & 2 & 3 & $2/3 \approx 0.666666$ & 6  \\
\hline
$B$ & 2 & 4 & $-1/2 = -0.5$ & 8 \\
\hline
$C_{1}$ & 0 & 7 & $0/7 = 0$ & 9-23 \\
\hline
$C_{2}$ & 0 & 7 & $0/7 = 0$ & 17 \\
\hline
$\bar{E}$ & 8 & 30 & $-4/15 \approx -0.266666$ & 21 \\
\hline
$F$ & 4 & 36 & $-1/9 \approx -0.111111$ & 15-29 \\
\hline
$G^{n}$ & 14 & 42 & $-1/3 \approx -0.333333$ & 24-38 \\
\hline
\end{tabular}
\caption{Mobile self-localizations properties such as solitons.}
\label{solitonGlidersTable}
\end{table}

Table~\ref{solitonGlidersTable} shows a number of properties to each mobile self-localization (Fig.~\ref{glidersSolitonR110}), such as: shift, period, speed, and volume (that can be related as its mass as well). All of them shall help us to synchronize collisions given a specific {\it phase}, where each mobile self-localization may present different {\it contact points} and collide with others mobile self-localizations. To produce a specific collision between mobile self-localizations at a given point we must have a full control over initial conditions, including distance between gliders and their phases at the moment of collision. 

The notation proposed to codify initial conditions in Rule 110 by phases is as follows:

\begin{equation}
\#_{1}(\#_{2},\mbox{f}_{i}\_1)
\end{equation}

\noindent where \#$_{1}$ represents a particular mobile self-localization (given in Table~\ref{solitonGlidersTable}) and \#$_{2}$ represents its phase if it has a period greater than four (for full details please see \cite{kn:MMS08}). Variable f$_{i}$ indicates the phase currently used where the second subscript $j$ (forming notation f$_{i}$\_$j$) indicates that selected master set of regular expressions.

In \cite{kn:MM01} we have calculated experimentally the whole set of binary collisions between mobile self-localizations in Rule 110, colliding all 1-1 mobile self-localizations. Thus in \cite{kn:Mart02a, kn:Mart02} we have reported all soliton reactions in Rule 110.

This way, 18 solitons (between two mobile self-localizations, i.e. binary) in Rule 110 can be coded in phases, as follows.

{\small
\begin{quote}
\begin{itemize}
\item[(a)] Soliton 1: $A$(f$_1\_1$)-$6e$-$G$(C,f$_1\_1$) $\longrightarrow$ $\{G,A\}$
\item[(b)] Soliton 2: $C_{1}$(A,f$_1\_1$)-$3e$-${\bar{E}}$(B,f$_1\_1$) $\longrightarrow$ $\{{\bar{E}},C_{1}\}$
\item[(c)] Soliton 3: $C_{1}$(A,f$_1\_1$)-$3e$-${\bar{E}}$(C,f$_1\_1$) $\longrightarrow$ $\{{\bar{E}},C_{1}\}$
\item[(d)] Soliton 4: $F$(A,f$_1\_1$)-$3e$-$B$(f$_4\_1$) $\longrightarrow$ $\{B,F\}$
\item[(e)] Soliton 5: $C_{2}$(A,f$_1\_1$)-$3e$-${\bar{E}}$(C,f$_1\_1$) $\longrightarrow$ $\{{\bar{E}},C_{2}\}$
\item[(f)] Soliton 6: $C_{1}$(A,f$_1\_1$)-$2e$-$F$(B,f$_1\_1$) $\longrightarrow$ $\{F,C_{1}\}$
\item[(g)] Soliton 7: $C_{2}$(A,f$_1\_1$)-$2e$-$F$(A,f$_1\_1$) $\longrightarrow$ $\{F,C_{2}\}$
\item[(h)] Soliton 8: $A$(f$_1\_1$)-$4e$-${\bar{E}}$(A,f$_1\_1$) $\longrightarrow$ $\{{\bar{E}},A\}$
\item[(i)] Soliton 9: $A$(f$_1\_1$)-$4e$-${\bar{E}}$(B,f$_1\_1$) $\longrightarrow$ $\{{\bar{E}},A\}$
\item[(j)] Soliton 10: $A$(f$_1\_1$)-$4e$-${\bar{E}}$(C,f$_1\_1$) $\longrightarrow$ $\{{\bar{E}},A\}$
\item[(k)] Soliton 11: $A$(f$_1\_1$)-$4e$-${\bar{E}}$(H,f$_1\_1$) $\longrightarrow$ $\{{\bar{E}},A\}$
\item[(l)] Soliton 12: $F$(A,f$_1\_1$)-$e$-${\bar{E}}$(A,f$_1\_1$) $\longrightarrow$ $\{{\bar{E}},F\}$
\item[(m)] Soliton 13: $F$(A,f$_1\_1$)-$e$-${\bar{E}}$(C,f$_1\_1$) $\longrightarrow$ $\{{\bar{E}},F\}$
\item[(n)] Soliton 14: $F$(A,f$_1\_1$)-$e$-${\bar{E}}$(D,f$_1\_1$) $\longrightarrow$ $\{{\bar{E}},F\}$
\item[(o)] Soliton 15: $F$(A,f$_1\_1$)-$e$-${\bar{E}}$(E,f$_1\_1$) $\longrightarrow$ $\{{\bar{E}},F\}$
\item[(p)] Soliton 16: $F$(G,f$_1\_1$)-$e$-${\bar{E}}$(A,f$_1\_1$) $\longrightarrow$ $\{{\bar{E}},F\}$
\item[(q)] Soliton 17: $F$(G,f$_1\_1$)-$e$-${\bar{E}}$(B,f$_1\_1$) $\longrightarrow$ $\{{\bar{E}},F\}$
\item[(r)] Soliton 18: $F$(G,f$_1\_1$)-$e$-${\bar{E}}$(H,f$_1\_1$) $\longrightarrow$ $\{{\bar{E}},F\}$
\end{itemize}
\end{quote}
}

\begin{figure}
\centerline{\includegraphics[width=4.2in]{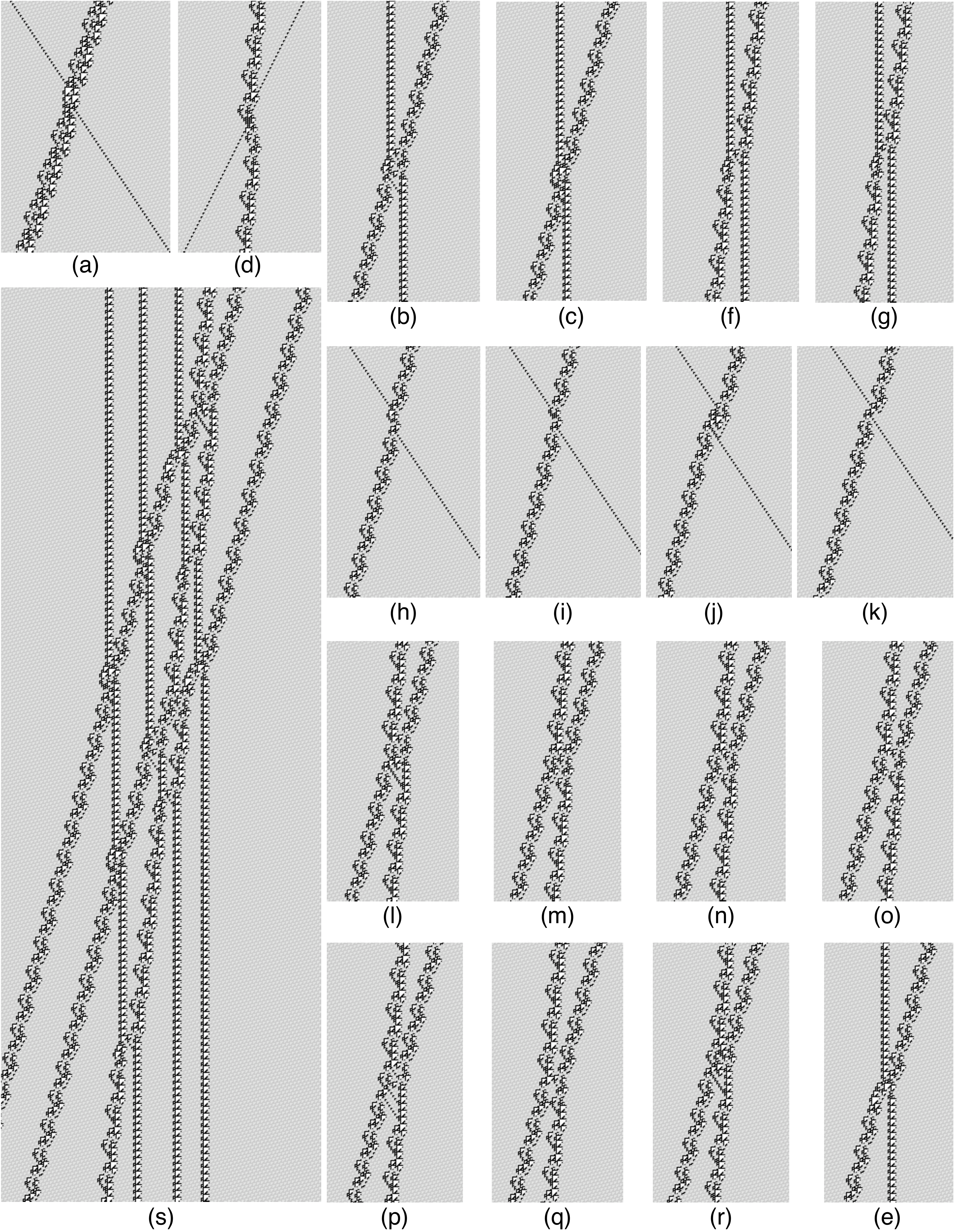}}
\caption{(a--r) Binary solitons in Rule 110, and one case (s) illustrating multiple solitonic collision with six mobile self-localizations, synchronized and evolving in  964 generations.}
\label{solitonsR110}
\end{figure}

\begin{figure}
\centerline{\includegraphics[width=2.3in]{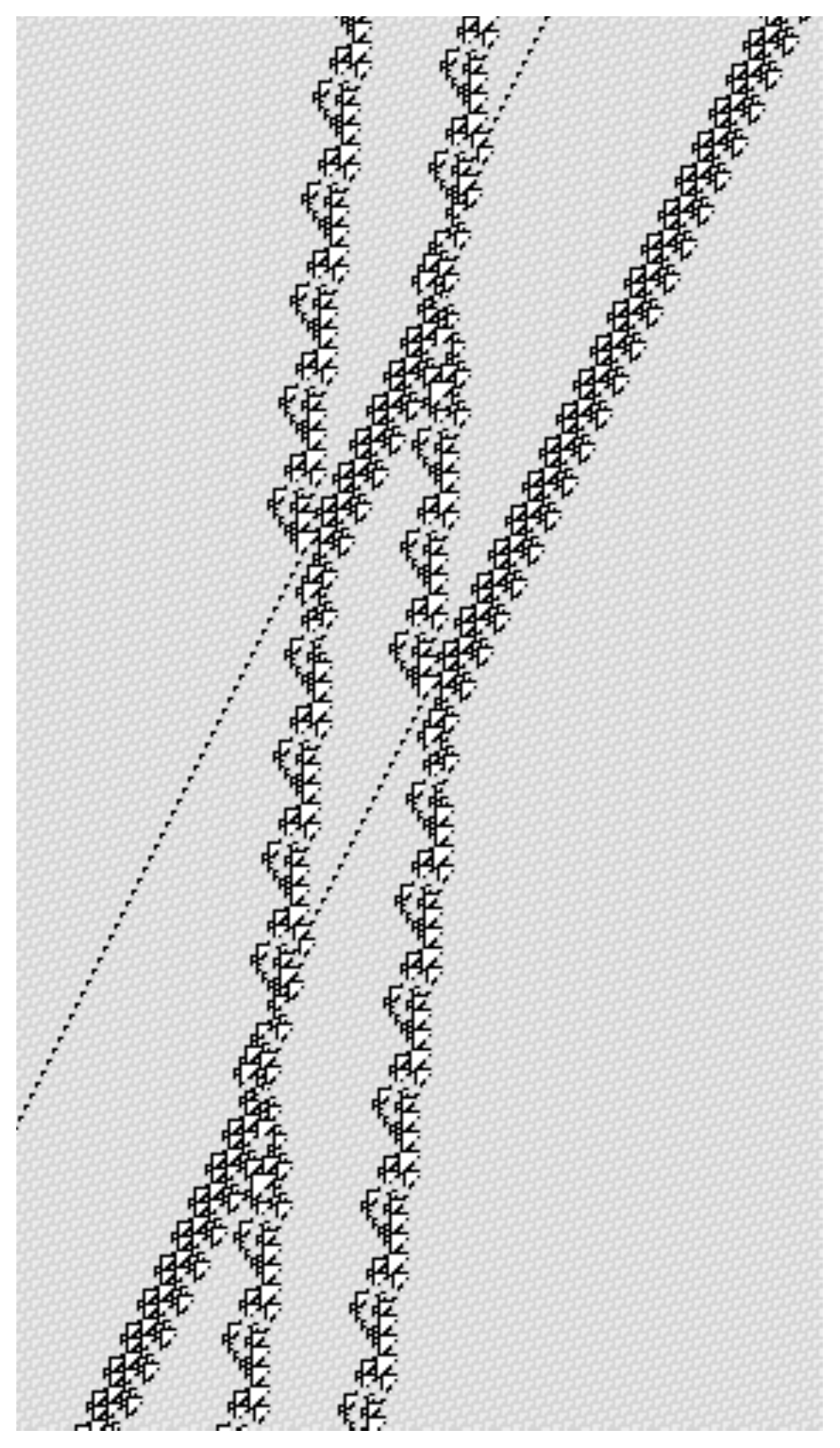}}
\caption{Pseudo-soliton in Rule 110.}
\label{pseudoSoliton}
\end{figure}

\begin{table}
\centering
\small
\begin{tabular}{l|l}
collisions $F \rightarrow {\bar{B}}$ & collisions $F \rightarrow B$ \\
\hline
$F$(A,f$_1\_1$)-$e$-${\bar{B}}$(A,f$_1\_1$) = $\{A, B, {\bar{B}}, F\}$ & $F$(A,f$_1\_1$)-$e$-$B$(f$_1\_1$) = $\{{\bar{B}}, F\}$ * \\
$F$(A,f$_1\_1$)-$e$-${\bar{B}}$(B,f$_1\_1$) = $\{A, 2C_3, C_1\}$ & $F$(G,f$_1\_1$)-$e$-$B$(f$_1\_1$) = $\{{\bar{B}}, F\}$ * \\
$F$(A,f$_1\_1$)-$e$-${\bar{B}}$(C,f$_1\_1$) = $\{A, C_2\}$ & $F$(H,f$_1\_1$)-$e$-$B$(f$_1\_1$) = $\{D_2, A^2\}$ \\
$F$(G,f$_1\_1$)-$e$-${\bar{B}}$(A,f$_1\_1$) = $\{C_2, A^2\}$ & $F$(A2)-$e$-$B$ = $\{B, F\}$ {\it (soliton)} \\
$F$(G,f$_1\_1$)-$e$-${\bar{B}}$(B,f$_1\_1$) = $\{A, A^3, A, {\bar{E}}$ & \\
$F$(G,f$_1\_1$)-$e$-${\bar{B}}$(C,f$_1\_1$) = $\{B, F\}$ * & \\
$F$(H,f$_1\_1$)-$e$-${\bar{B}}$(A,f$_1\_1$) = $\{A, C_2\}$ & \\
$F$(H,f$_1\_1$)-$e$-${\bar{B}}$(B,f$_1\_1$) = $\{{\bar{E}, A^5}\}$ & \\
$F$(H,f$_1\_1$)-$e$-${\bar{B}}$(C,f$_1\_1$) = $\{{\bar{E}, A^5}\}$ & \\
$F$(A2,f$_1\_1$)-$e$-${\bar{B}}$(A,f$_1\_1$) = $\{C_1\}$ & \\
$F$(A2,f$_1\_1$)-$e$-${\bar{B}}$(B,f$_1\_1$) = $\{A, B^3, {\bar{E}}\}$ &
\end{tabular}
\caption{Reactions relation between $B$, ${\bar{B}}$ and $F$ mobile slef- localizations in Rule 110.}
\label{pseudoSolitonTable}
\end{table}

Of course, from these solitonic binary collisions we can codify and synchronize most structures and therefore to get multiple solitonic reactions increasing its complexity. For example, we have the next codification:

{\small
\begin{quote}
\begin{itemize}
\item[(s)] Multiple soliton:  $C_1$(B,f$_1\_1$)-$e$-$C_1$(A,f$_1\_1$)-$2e$-$C_2$(A,f$_1\_1$)-$e$-$F$(A,f$_1\_1$)-$e$-${\bar{E}}$(A,f$_1\_1$)-$3e$-${\bar{E}}$(C,f$_2\_1$) $\longrightarrow$ $\{{\bar{E}},{\bar{E}},F,C_1,C_1,C_2\}$.
\end{itemize}
\end{quote}
}

All these solitons in Rule 110 are displayed in Fig.~\ref{solitonsR110}. Each codification of (a) to (r) present the binary case, and (s) presents a multiple solitonic collision with six localizations, where each is synchronized to produce the soliton reaction.

Yet as an special case in Rule 110, we can find a collision named as {\it pseudo-soliton} \cite{kn:Mart02a}, that works recovering the original localization after two collisions. This is performed with $B$, ${\bar{B}}$ and $F$ localizations. Localizations $B$ and ${\bar{B}}$ have the same period and speed, but its volume is different.

From ``Atlas of binary collisions in Rule 110'' \cite{kn:MM01} we have calculated the whole set of binary collisions between mobile self-localizations, and summarised them in Table~\ref{pseudoSolitonTable}. We have placed particular attention on asterisks labels because they represent precisely the pseudo-soliton in Rule 110. Hence we know that reaction $F \rightarrow {\bar{B}} = \{B, F\}$ and also that $F \rightarrow B = \{{\bar{B}}, F\}$, thus a loop may be constructed to synchronize such collisions. Figure~\ref{pseudoSoliton} displays such construction, given its codification in phases as:

{\small
\begin{center}
$F$(G,f$_{3}\_1$)-$2e$-$F$(A,f$_{1}\_1$)-$e$-$B$(f$_{1}\_1$)-$5e$-$\bar{B}$(B,f$_{4}\_1$).
\end{center}
}

\subsection{Solitons in ECA Rule 54}
\label{54-solitons}

ECA Rule 54 is a complex CA evolving with an ``apparently'' simple system of mobile self-localizations. Its local function is defined as follows:

\begin{equation}
\varphi_{R54} = \left\{
	\begin{array}{lcl}
		1 & \mbox{if} & 001, 010, 100, 101 \\
		0 & \mbox{if} & 000, 011, 110, 111
	\end{array} \right. .
\end{equation}

\begin{figure}[th]
\centerline{\includegraphics[width=4.2in]{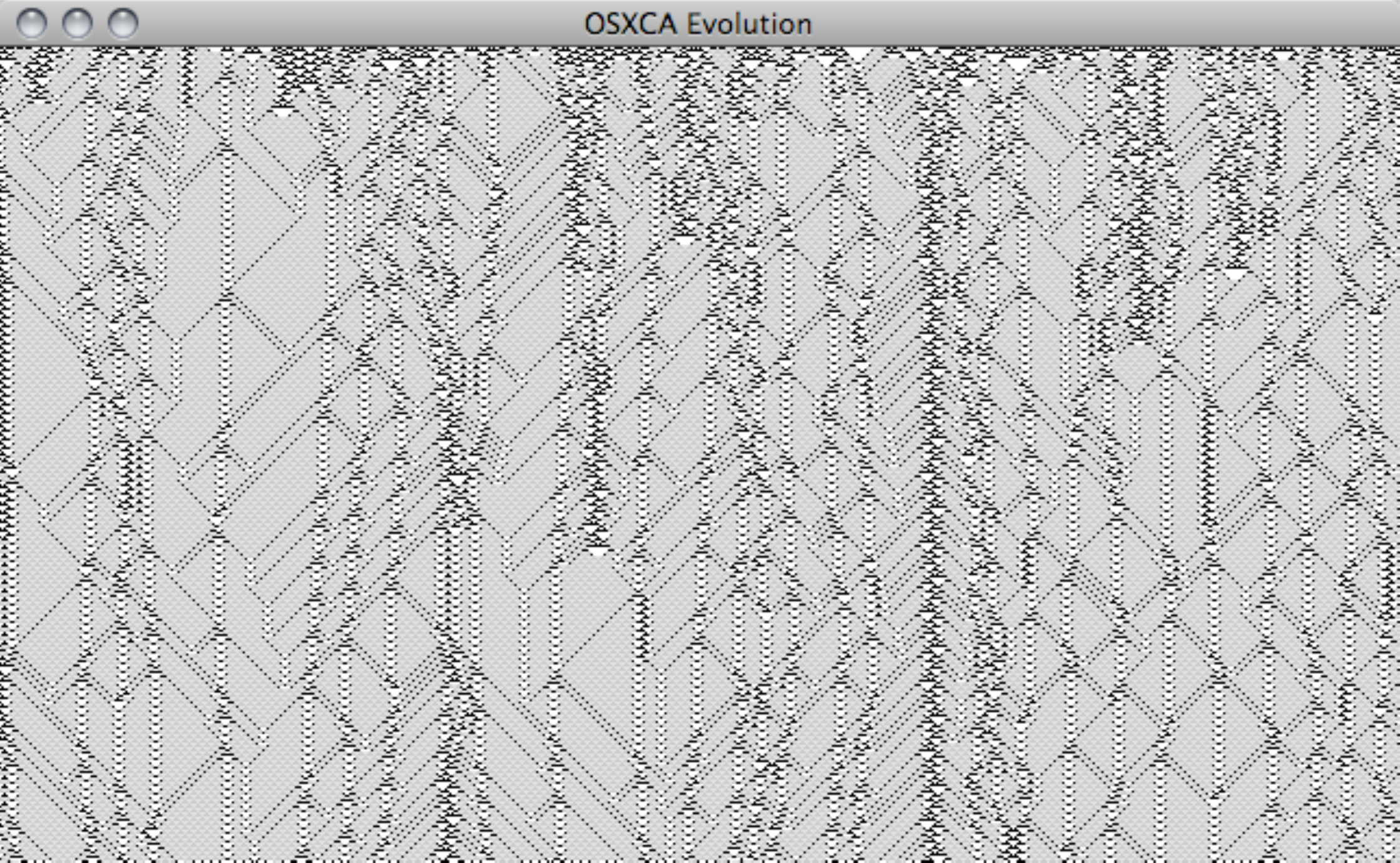}}
\caption{Random evolution in Rule 54 on a ring of 644 cells to 375 generations. White cells represent the state 0 and black cells the state 1, the evolution start with a density at 50\%. Also a filter is applied to get a better view of mobile self-localizations and collisions on its periodic background.}
\label{evolR54}
\end{figure}

Figure~\ref{evolR54} illustrates the complex dynamics from a typical random initial condition selecting the evolution rule $\varphi_{R54}$. Here we can see how a number of mobile self-localizations emerge on its evolution space and collide. Particularity, Rule 54 is able to evolve with emergence of glider guns\footnote{A {\it glider gun} is a complex structure that emits periodically a localization, famously known in the Game of Life CA.} patterns since random initial conditions while this fact is not common in Rule 110 or from others ECA rules.

Detailed analysis of various aspects of ECA Rule 54 can be found in: localizations system\footnote{Gliders in Rule 54 \url{http://uncomp.uwe.ac.uk/genaro/rule54/glidersRule54.html}.} \cite{kn:BNR91, kn:HC97, kn:MAM06, kn:MAM08}, computations \cite{kn:MAM06}, collisions \cite{kn:MAM06}, algebraic properties \cite{kn:Red10, kn:Mar00}. So generalities can be explored from the {\it Rule 54 repository}.\footnote{Rule 54 repository: \url{http://uncomp.uwe.ac.uk/genaro/Rule54.html}.}

\begin{figure}[th]
\centerline{\includegraphics[width=4in]{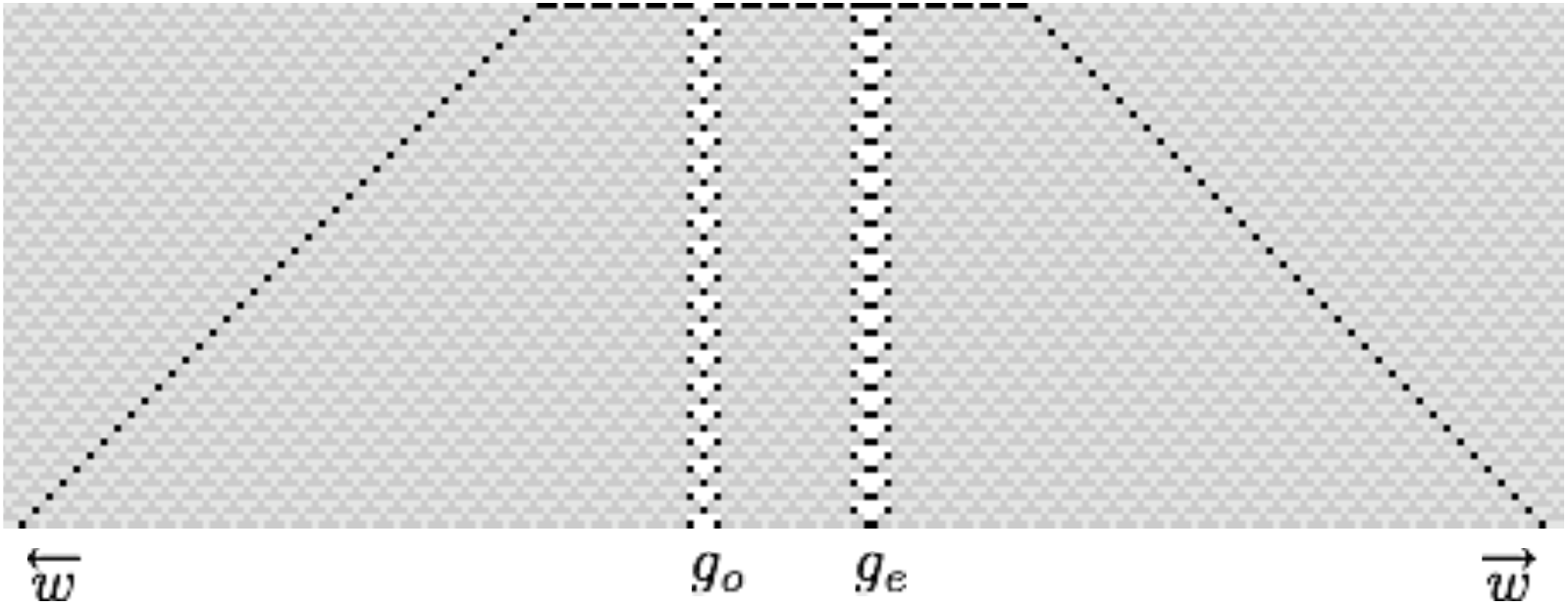}}
\caption{Set of mobile self-localizations with solitonic properties in Rule 54.}
\label{glidersSolitonR54}
\end{figure}

The set of mobile self-localizations in Rule 54 is significantly small compared with Rule 110 (that has a base set of 12 mobile self-localizations). Rule 54 has basically four primitive or basic  mobile self-localizations (stationary, shift-right, and shift-left displacements) and three kinds of glider guns \cite{kn:MAM06}. This way, basic mobile self-localizations work to produce solitons from multiple collisions. In Fig.~\ref{glidersSolitonR54} we present these basic mobile self-localizations following the Boccara's notation \cite{kn:BNR91}.

\begin{figure}
\centerline{\includegraphics[width=4.2in]{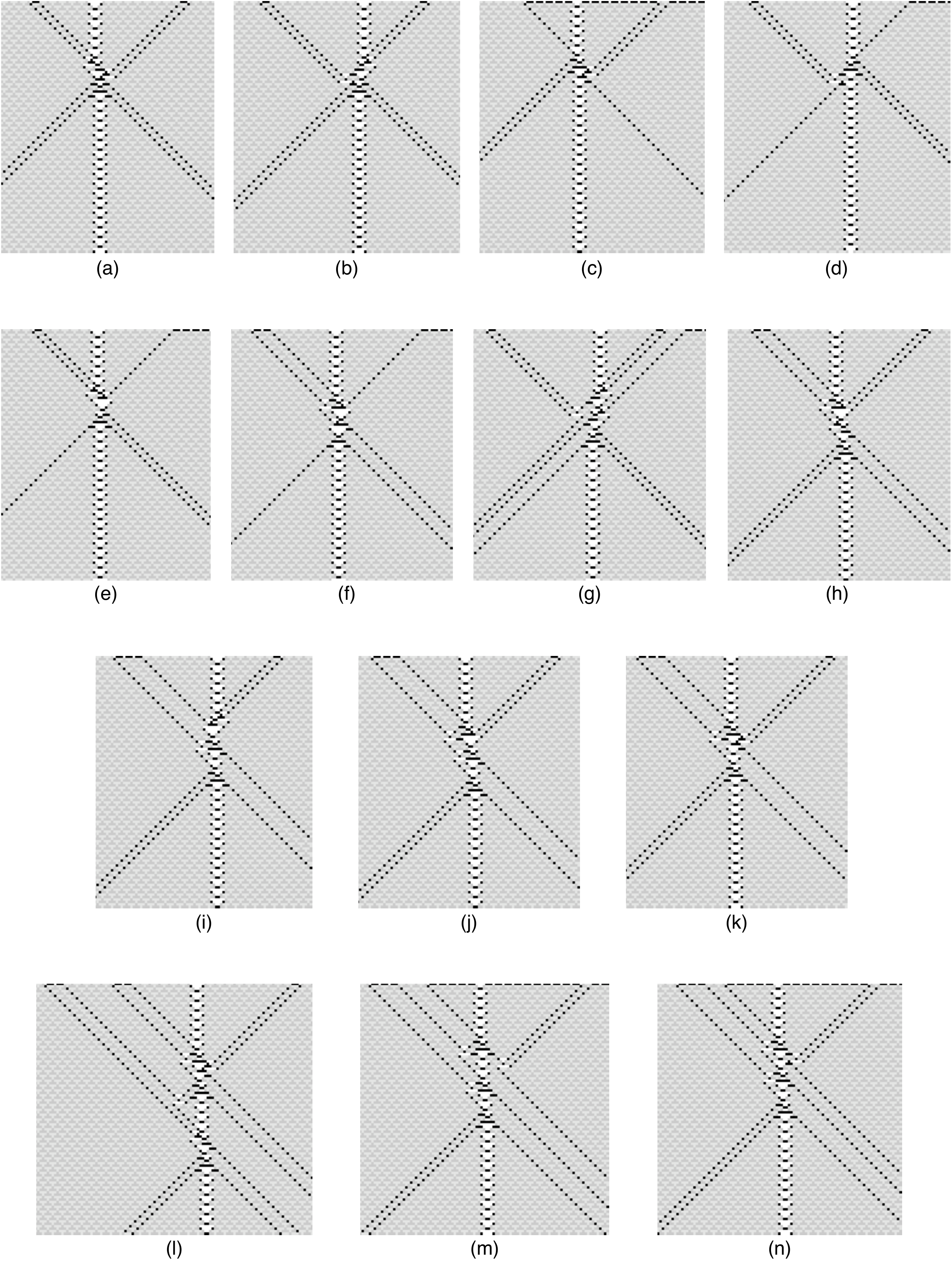}}
\caption{Cataloge of soliton collisions in Rule 54.}
\label{solitonsR54}
\end{figure}

Particularly, solitons in Rule 54 cannot emerge from binary collisions, they are found in multiple collisions. This way, solitons there are on the domain of triple collisions and beyond \cite{kn:MAM06}. Properties for these mobile self-localizations are characterized in Table~\ref{solitonGlidersR54Table}.

In this paper, we will report an unexplored set of solitonic collisions in Rule 54. They are obtained for systematic analysis by reactions across of multiple collisions.

\begin{table}[th]
\centering
\small
\begin{tabular}{|c|c|c|c|c|}
\hline
glider & shift & period & speed & volume \\
\hline \hline
$\overrightarrow{w}$ & 2 & 2 & 1 & 2 \\
\hline
$\overleftarrow{w}$ & $-2$ & 2 & $-1$ & 0-4 \\
\hline
$g_{o}$ & 0 & 4 & 0 & 6-2 \\
\hline
$g_{e}$ & 0 & 4 & 0 & 7-3 \\
\hline
\end{tabular}
\caption{Basic mobile self-localizations properties in Rule 54.}
\label{solitonGlidersR54Table}
\end{table}

This way, Fig.~\ref{solitonsR54} presents 14 kinds of solitons constructed in Rule 54. It is easy to recognise that you can derive 28 similar reactions in total, because Rule 54 is a symmetric rule and therefore you can obtain the next 14 symmetric collisions.

We can see that Fig. (a) and (b) display two pairs of mobile self-localizations producing the same soliton reaction, however the collision is different because while in (a) the first pair of mobile self-localizations delay its trajectory in (b) it advances for six cells. Thus here is possible controller intervals of mobile self-localizations trajectories. Similar cases are presented in Fig. (c) and (d), but these reactions are between three mobile self-localizations.

Figure (g) starts solitonic reaction with more than four mobile self-localizations, but here we employ more space between intervals of mobile self-localizations (see (f) and (h) as well). Hence we can use several mobile self-localizations to preserve the soliton reaction. Noticeably, Fig. (i), (j) and (k) display three different kinds of collisions to get soliton reactions employing four mobile self-localizations.

The last set of collisions (Fig. (l), (m), and (n)) display more large synchronisations of mobile self-localizations, with different intervals and numbers of them. Of course, it is possible to design more sophisticated collisions working with a diversity of packages of mobile and stationary  self-localizations.

\begin{figure}[th]
\centerline{\includegraphics[width=2.5in]{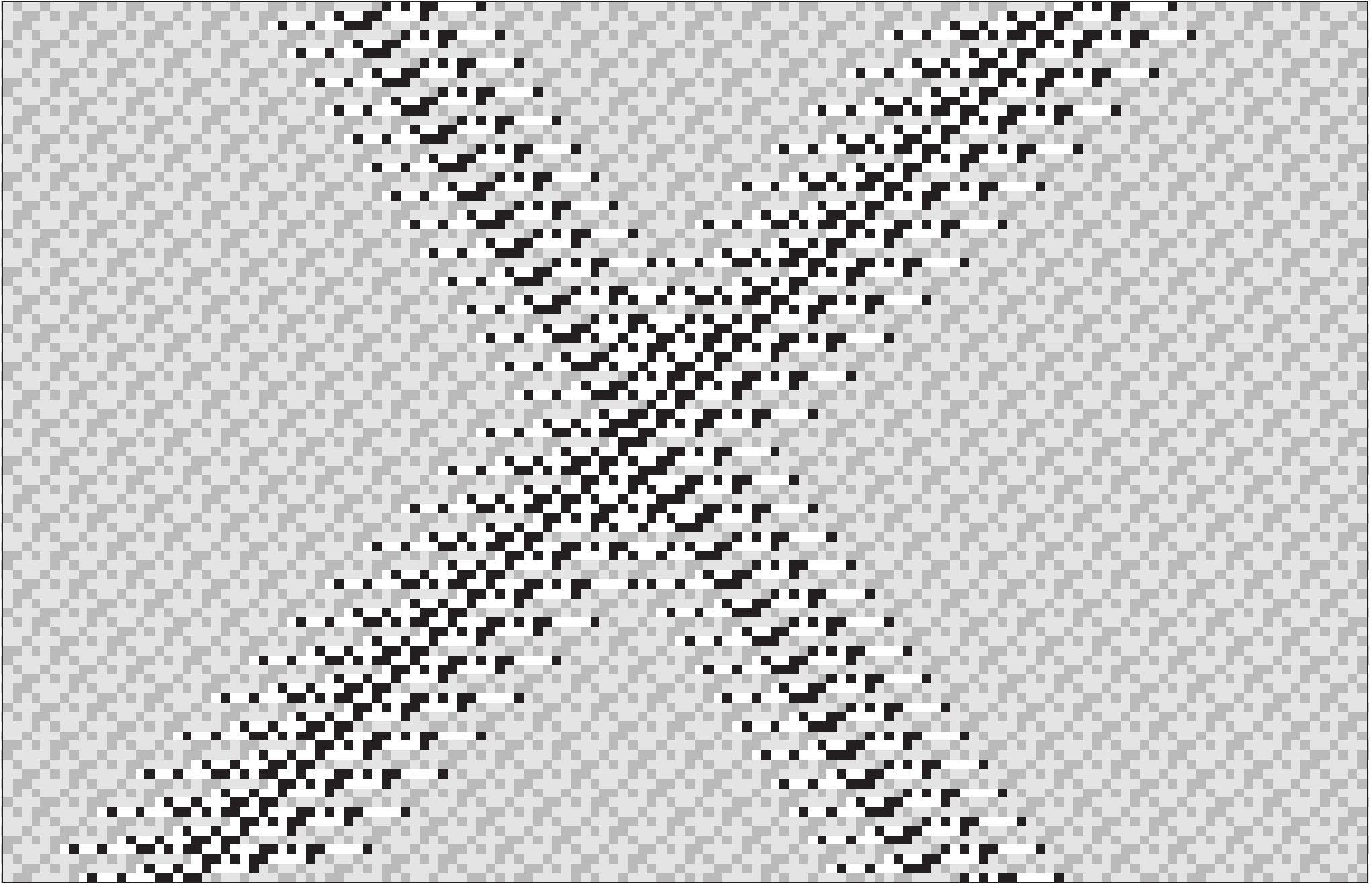}}
\caption{Soliton in Class II ECA Rule 26.}
\label{solitonECAr26}
\end{figure}

By the way, recently a soliton is discovered in ECA Rule 26 (Fig.~\ref{solitonECAr26}).

\subsection{Solitons in ECAM}
\label{Mem-solitons}

In this section, we will present the simple single-soliton two-component solution \cite{kn:Steig00} for a specific ECAM. The main characteristic is that only one mobile self-localization is processed. Thus a mobile self-localization with shift-right and shift-left displacement always produce the same reaction.

Actually an extensive and systematic analysis is done for ECAM, in ``Designing Complex Dynamics with Memory'' \cite{kn:MAA}. From here, we have selected the ECAM rule $\phi_{R9maj:4}$, because none another rule have the same features.

\begin{figure}[th]
\centering
\subfigure[]{\scalebox{0.32}{\includegraphics{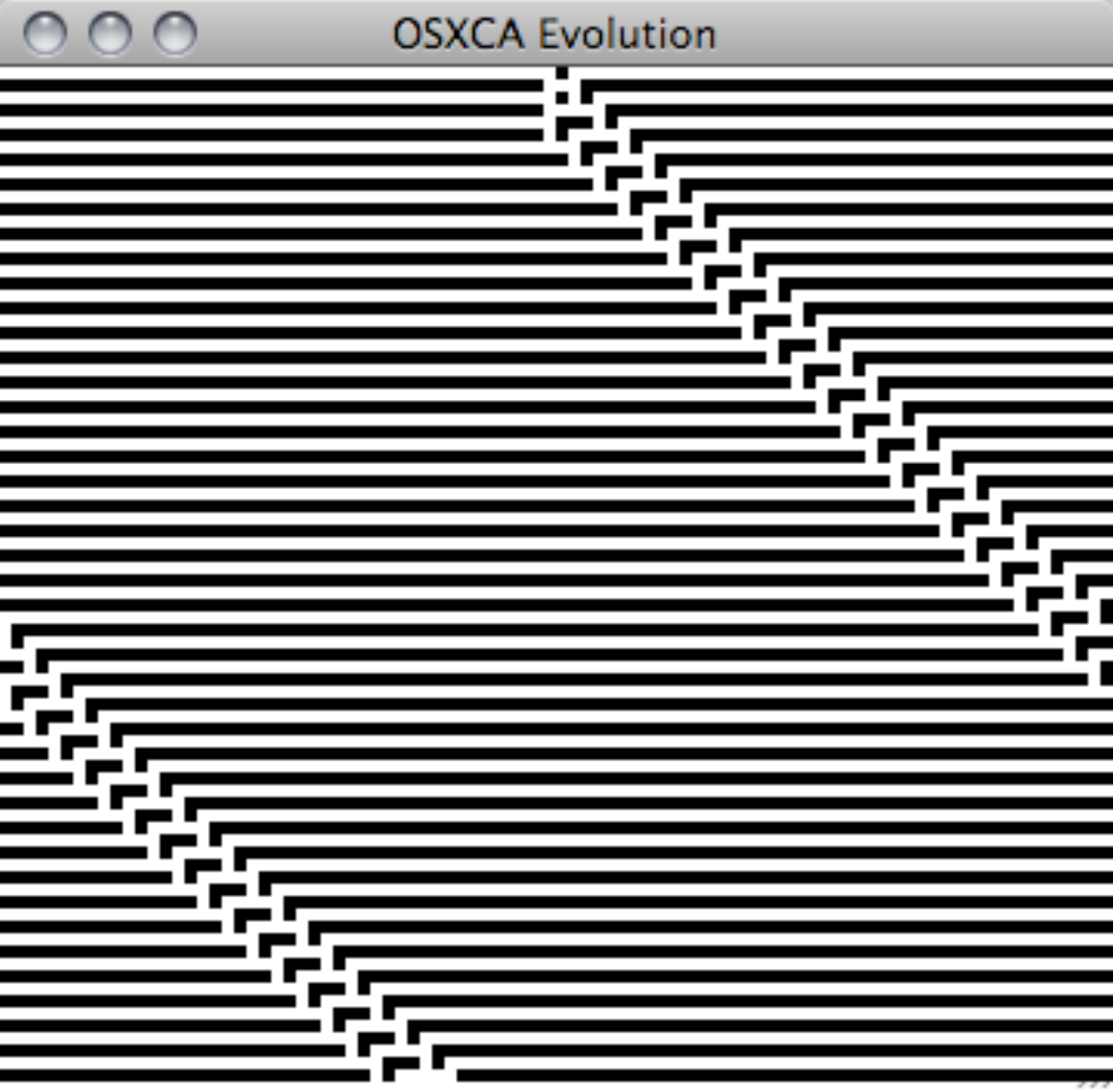}}} \hspace{0.8cm}
\subfigure[]{\scalebox{0.32}{\includegraphics{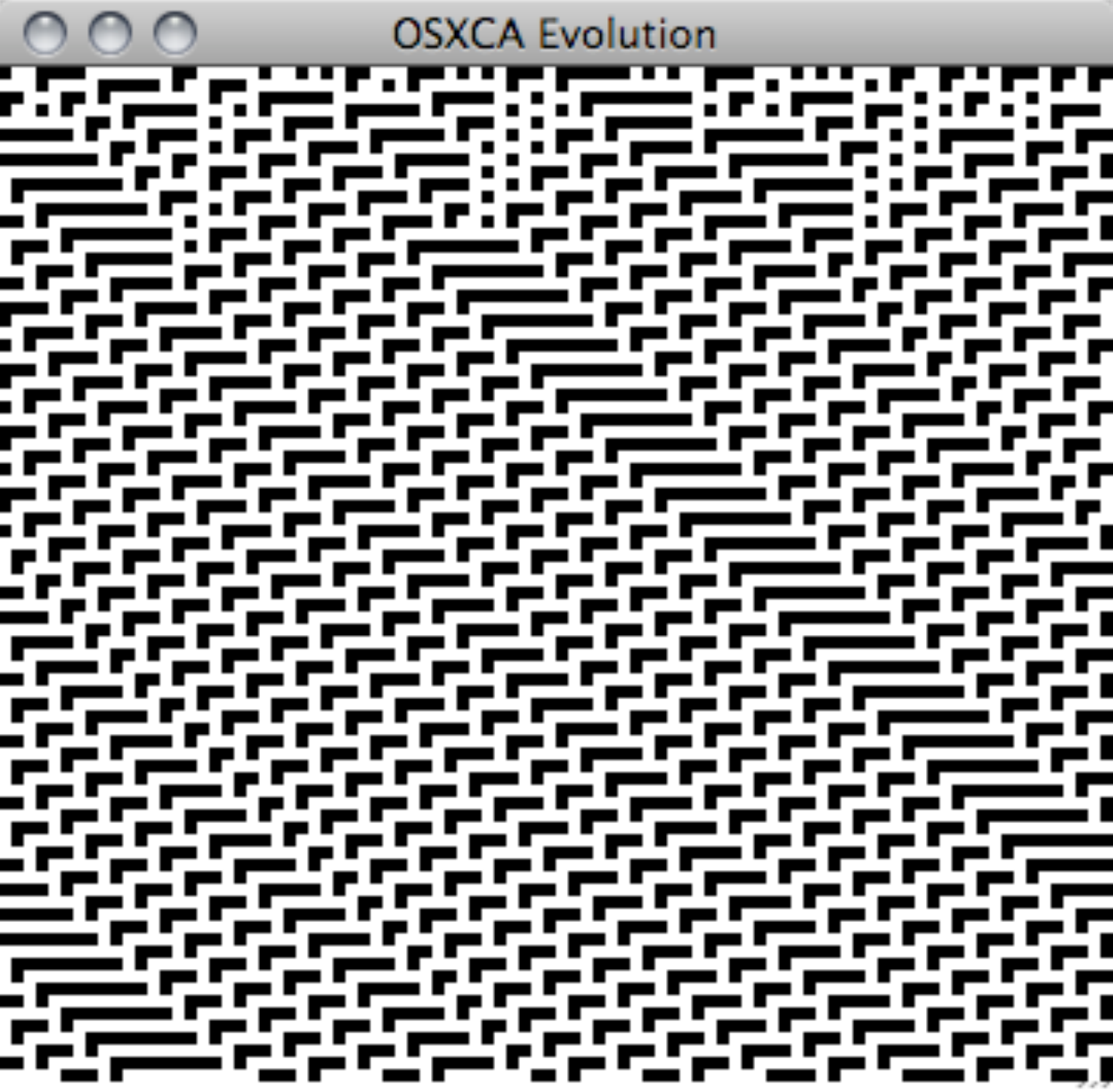}}}  
\subfigure[]{\scalebox{0.5}{\includegraphics{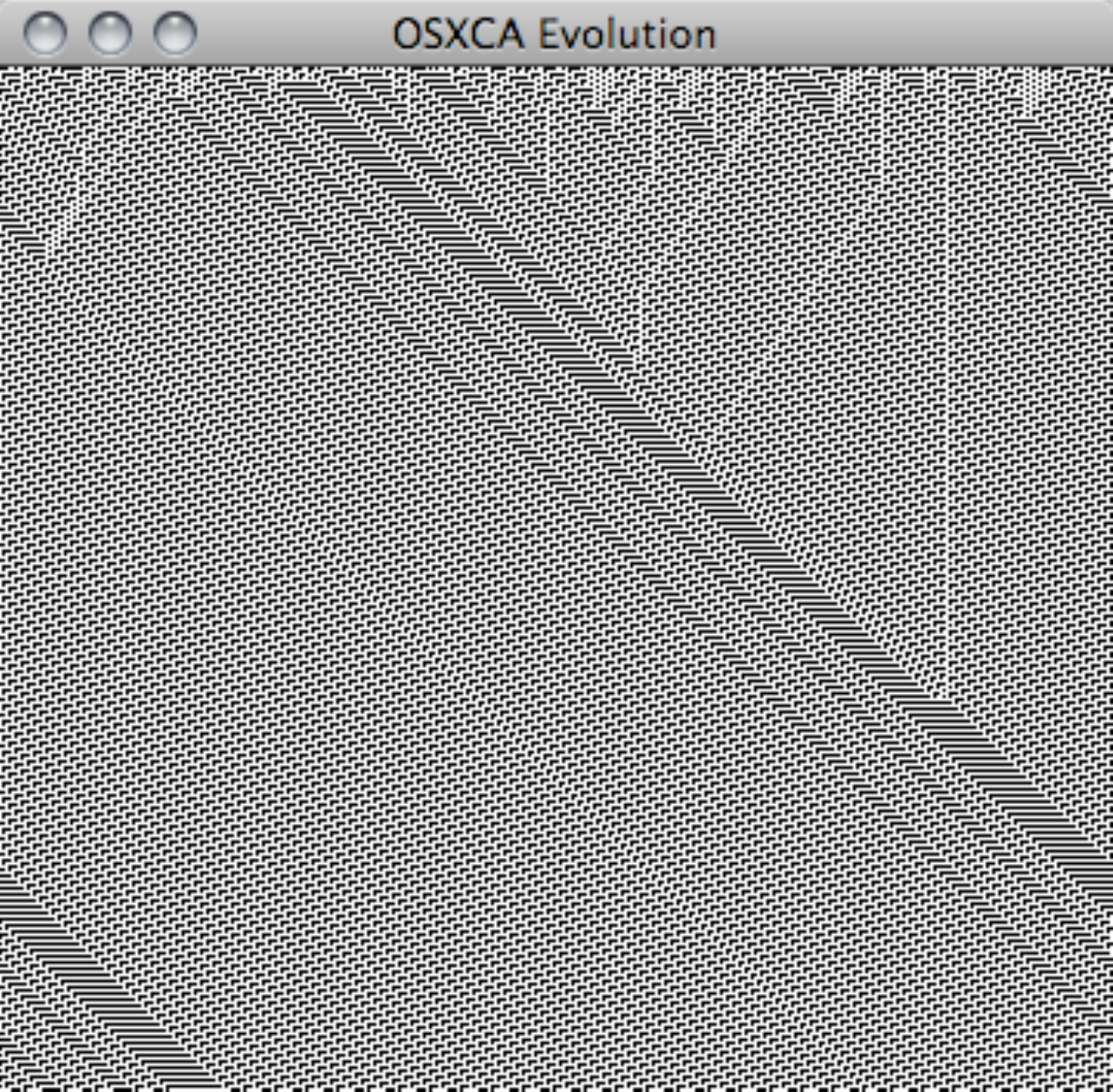}}} 
\caption{Typical snapshots of ECA Rule 9. (a) starts evolution with a single cell in state one, (b) presents a random evolution at 50\%, and (c) other random evolution with small pixels, 360 cells for 331 generations.}
\label{ECAR9}
\end{figure}

\begin{figure}
\centering
\subfigure[]{\scalebox{0.385}{\includegraphics{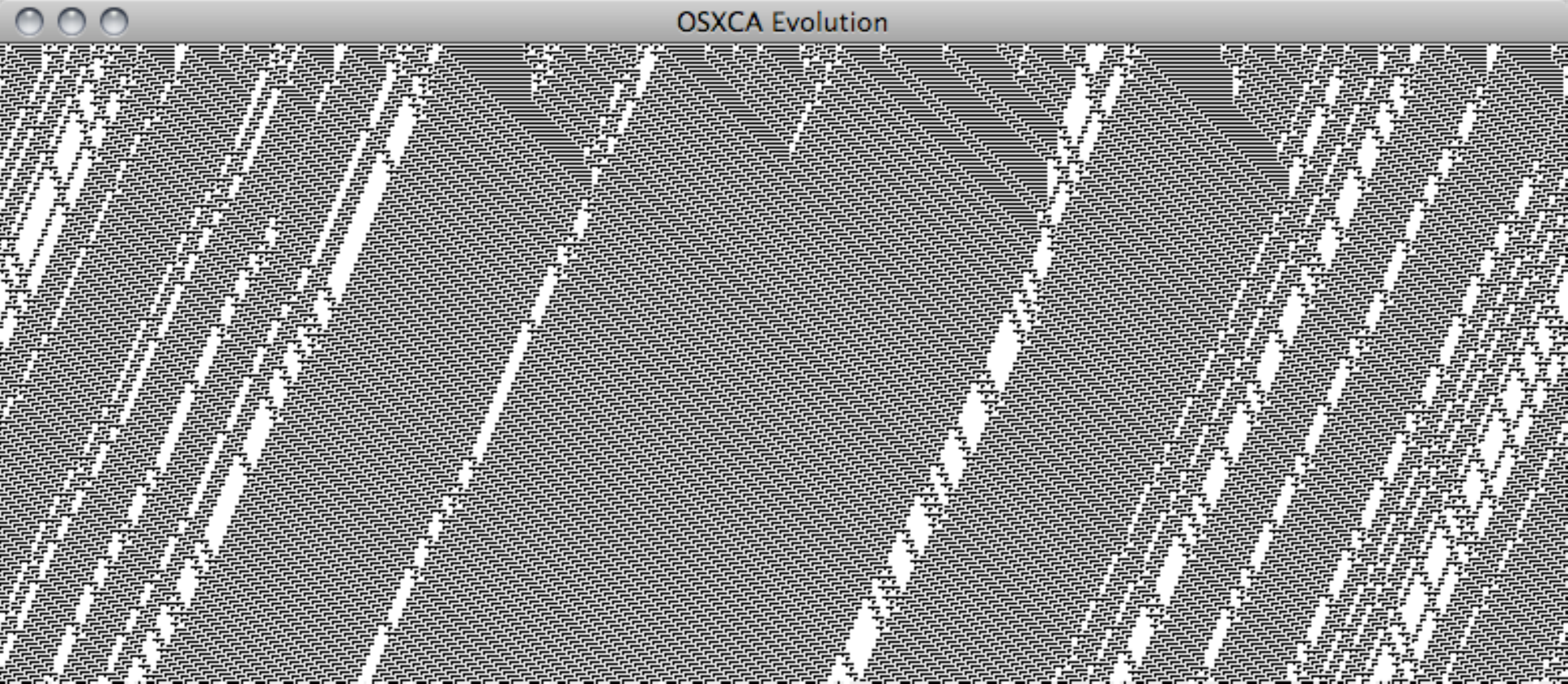}}} 
\subfigure[]{\scalebox{0.385}{\includegraphics{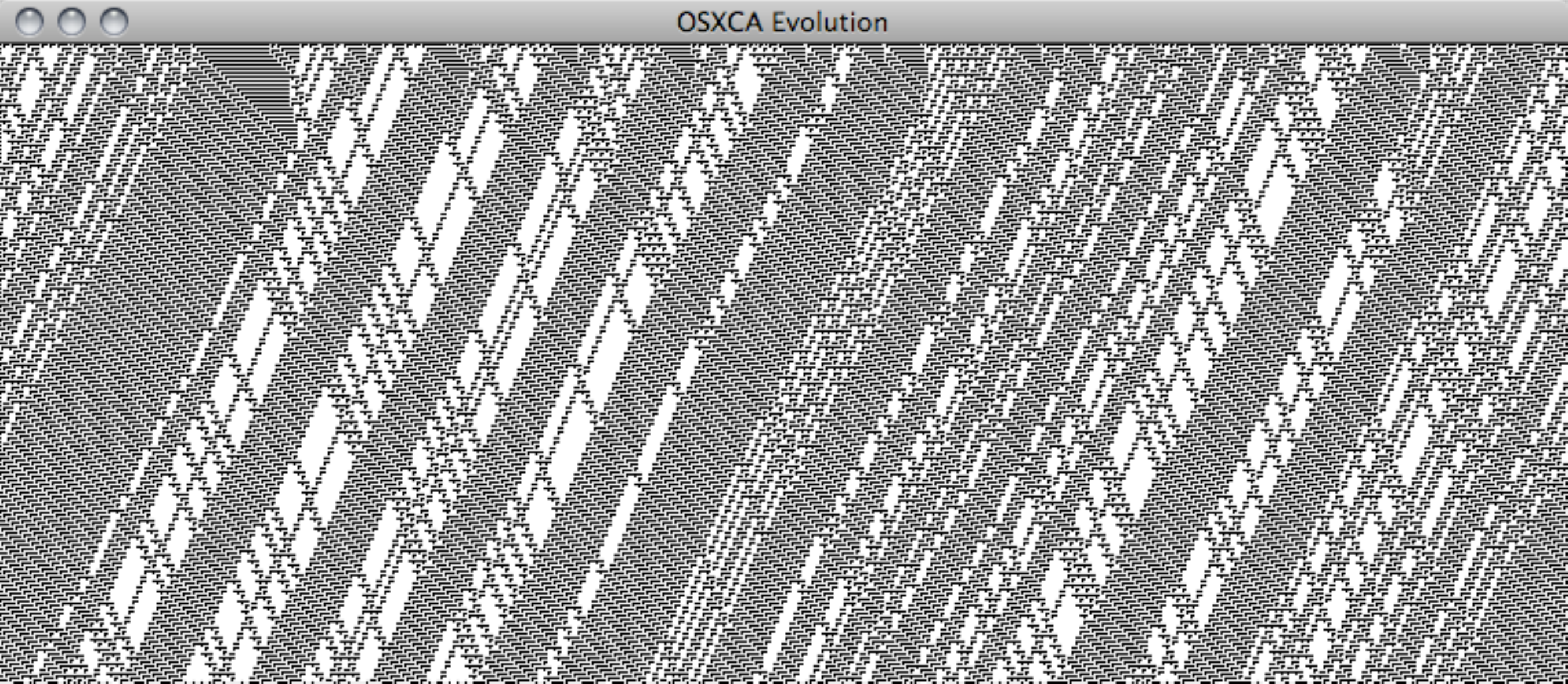}}}  
\subfigure[]{\scalebox{0.385}{\includegraphics{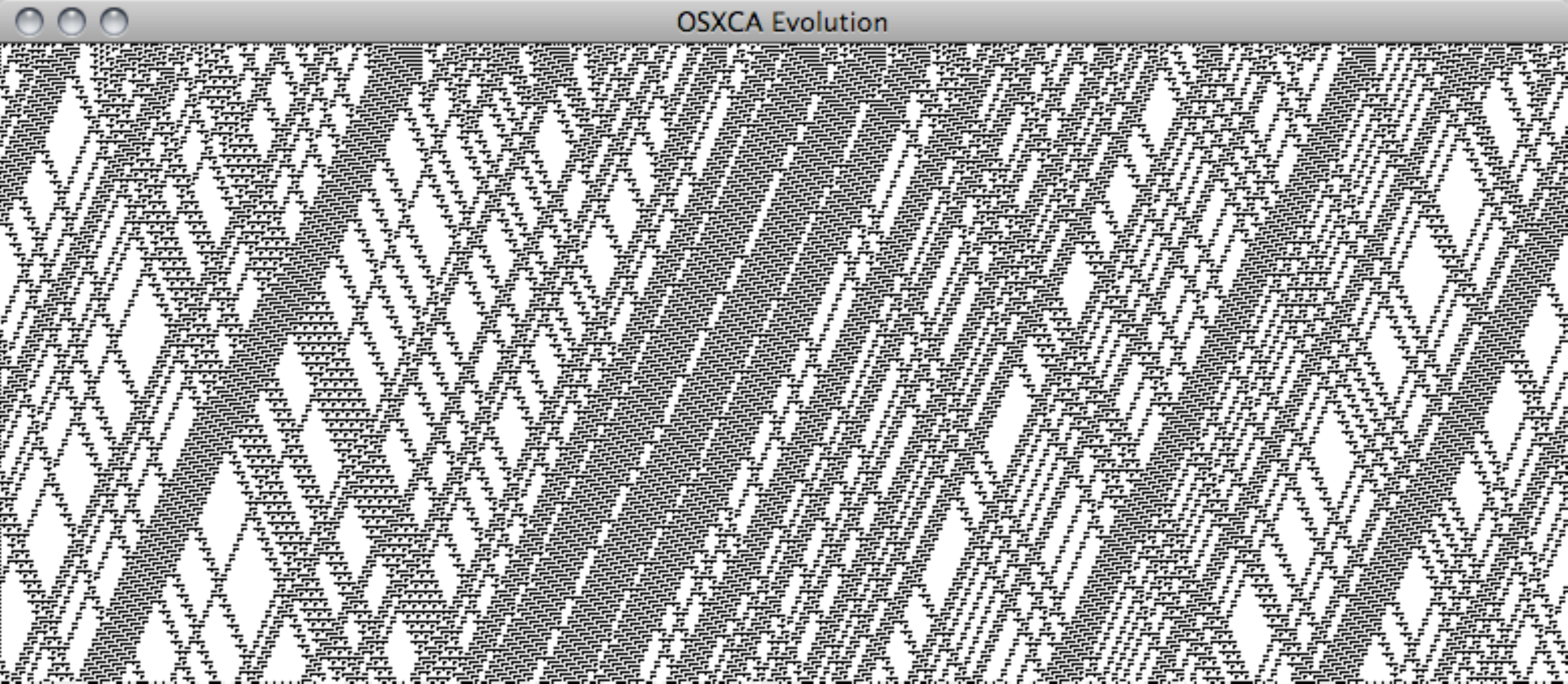}}} 
\caption{Typical snapshots of ECAM rule $\phi_{R9maj:4}$. (a) starts with an initial density of 10\%, (b) presents an initial density of 80\%, and (c) has an initial density of 50\%. All evolutions are filtered for the best visualization of mobile self-localization interaction, the evolutions are on a ring of 776 cells for 315 generations.}
\label{ECAMR9maj4}
\end{figure}

Figure~\ref{ECAR9} shows the ECA base that shall be enriched with {\it majority memory function} (see Sec.~\ref{ecam}). We study ECA Rule 9 because it displays basic interaction of solitons with simple collisions. As we can see in (a) a single soliton travels along the evolution space, while in (b) we can see a number of interactions during a short history, starting from a random initial condition. Extending the evolution space in (c) we can observe better how these solitons emerge in ECA Rule 9 and how they collide inside a fast stationary periodic attractor.

In \cite{kn:MAA10, kn:MAA12, kn:MAS10}, we have demonstrated how ECA when enriched with memory with memory produces different dynamics. Here we will exploit this tool to get simple solitonic reactions.

Let use the majority memory with $\tau=4$ in ECA Rule 9. Obtaining the  ECAM rule $\phi_{R9maj:4}$, that evolves with two mobile self-localizations emerging on its evolution space: $\cal G$$_{\phi_{R9maj:4}} = \{ \overrightarrow{p}, \overleftarrow{p} \}$. The localization's properties are easy to calculate. The $\overrightarrow{p}$ mobile self-localization has a volume of $5 \times 6$ cells, a mass of 12 cells, and moves 2 cells in 5 generations (shift-right displacement). While $\overleftarrow{p}$ mobile self-localization has a volume of $5 \times 3$ cells, a mass of 7 cells, and moves 2 cells in 5 generations (shift-left displacement).

Mobile self-localizations emerging in ECAM $\phi_{R9maj:4}$ preserve the {\it solitonic reaction} since any collision. Well, here really there are two different collisions (two contact points \cite{kn:PST86} or phases \cite{kn:MMS08} in every mobile self-localization) between $\overrightarrow{p}$ and $\overleftarrow{p}$ mobile self-localizations, but at the first collision the soliton is preserved because the sequence is fused in a string of four cells in state one, while the second reaction fuse a string of eight cells in state one. Although finally they open in both mobile self-localizations again late of exactly seven generations.

Thus the automaton $\phi_{R9maj:4}$ adjusts every string to always evolve with the same mobile self-localization and soliton reactions, as follows the next relation of collisions:

$$
\overrightarrow{p} \rightarrow \overleftarrow{p} = \{\overleftarrow{p}, \overrightarrow{p}\}, \mbox{ and }$$ $$
\overrightarrow{p} \leftarrow \overleftarrow{p} = \{\overleftarrow{p}, \overrightarrow{p}\}. 
$$

Figure~\ref{ECAMR9maj4} illustrates three different random initial conditions where the ECAM rule $\phi_{R9maj:4}$ always evolve in solitonic collisions. First evolution (Fig.~\ref{ECAMR9maj4}a) starts with an initial density of 10\% for state one, the result implies a high production of $\overleftarrow{p}$ mobile self-localizations with very few $\overrightarrow{p}$ mobile self-localizations, preserving always the solitonic collisions inside bigger fields of $\overleftarrow{p}$. In the opposite case, the second evolution (Fig.~\ref{ECAMR9maj4}b) has an initial density of 80\% for state one and produce again, high concentrations of $\overleftarrow{p}$ mobile self-localizations with some $\overrightarrow{p}$ mobile self-localizations but newly the solitonic reaction is always preserved. The final evolution (Fig.~\ref{ECAMR9maj4}c) displays a 50\% of states one and zero, generating a similar distribution of both mobile self-localizations. In all cases, the ECAM rule $\phi_{R9maj:4}$ evolve any initial condition in solitons. Thus you can begin with any number of mobile self-localizations in $\phi_{R9maj:4}$ and the solitons are always produced. A characteristic that no conventional ECA have. However, we will mention that such behaviour can be reproduced identically in other kinds of CA, the reversible {\it block CA} (or also known as partitioned 1D CA as well) explored by Wolfram in \cite{kn:Wolf02} (chapter 9).

\subsection{Computing with CA solitons}
\label{computing}

\begin{figure}[th]
\centerline{\includegraphics[width=4.2in]{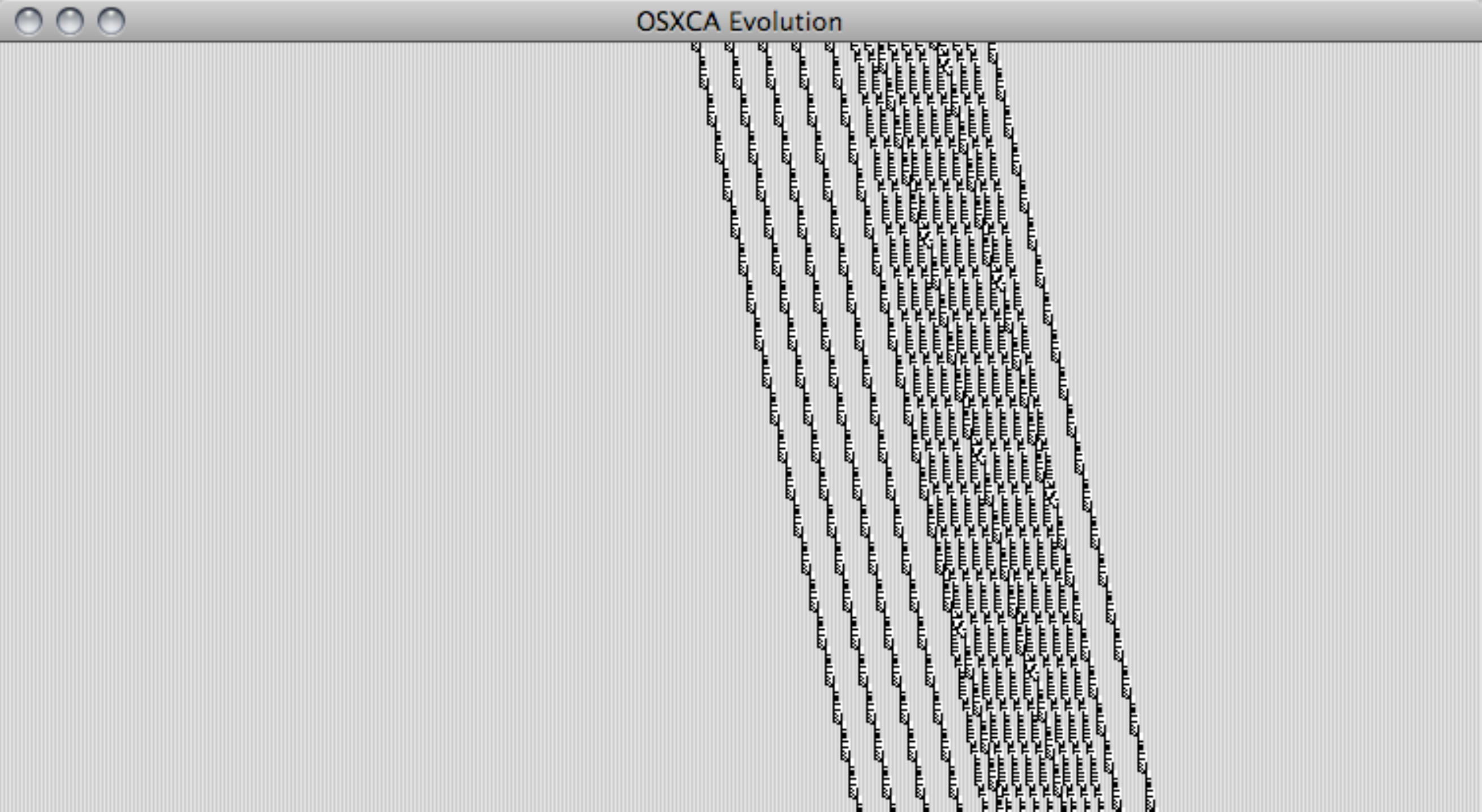}}
\caption{The ECAM $\phi_{R30maj:8}$ presents a solitonic collision that can be coded for any $n, m \in \mathcal{Z}^+$, such that, $p^n_{\phi_{R30maj:8}} \rightarrow q^m_{\phi_{R30maj:8}}$ is always a soliton.} 
\label{temporal}
\end{figure}

Solitons are useful to preserve information such as in the fibre-optic communications field. A particular interest is known if such solitons could emulate an equivalent Turing machine. Steiglitz {\it et. al} have designed a number of results trying to reach this goal, please see \cite{kn:JSS01, kn:RSP05, kn:SKW88, kn:Steig00}, and logic gates with solitons in \cite{kn:Ada02a, kn:BW02}. In \cite{kn:MAA12} for example, authors have developed a very simple substitution system as an implementation of the function {\sf addToHead()}, based in soliton reactions, where also such mechanisms can be designed as a simple collider \cite{kn:MAS11}. Figure~\ref{temporal} displays such operations between two mobile self-localizations in the ECAM $\phi_{R30maj:8}$ \cite{kn:MAA10, kn:MAA12}.

During the last decade we have seen a number of significant advances in work with solitons for modelling unconventional computing devices, you can see the next references \cite{kn:Ada01, kn:Ada02, kn:Ada02a, kn:BW02, kn:JSS01, kn:MAA12, kn:RSP05, kn:Siw02, kn:Steig00}. As a result, we can see how solitons could be important to develop computable devices in the construction of equivalent Turing machines. We also want to recall the results obtained in ECA Rule 110, where a cyclic tag system was developed to perform a computation based-collisions with a large number of mobile self-localizations on an incredible global synchronization in millions of cells. Solitonic reactions were very useful to write binary data and preserve information in the whole mechanism. For full details please see \cite{kn:Cook04, kn:Cook08, kn:Mc02, kn:MMS11, kn:Wolf02}.\footnote{Details and large snapshots about cyclic tag system working in Rule 110 \url{http://uncomp.uwe.ac.uk/genaro/rule110/ctsRule110.html}}

\begin{figure}
\centerline{\includegraphics[width=4.2in]{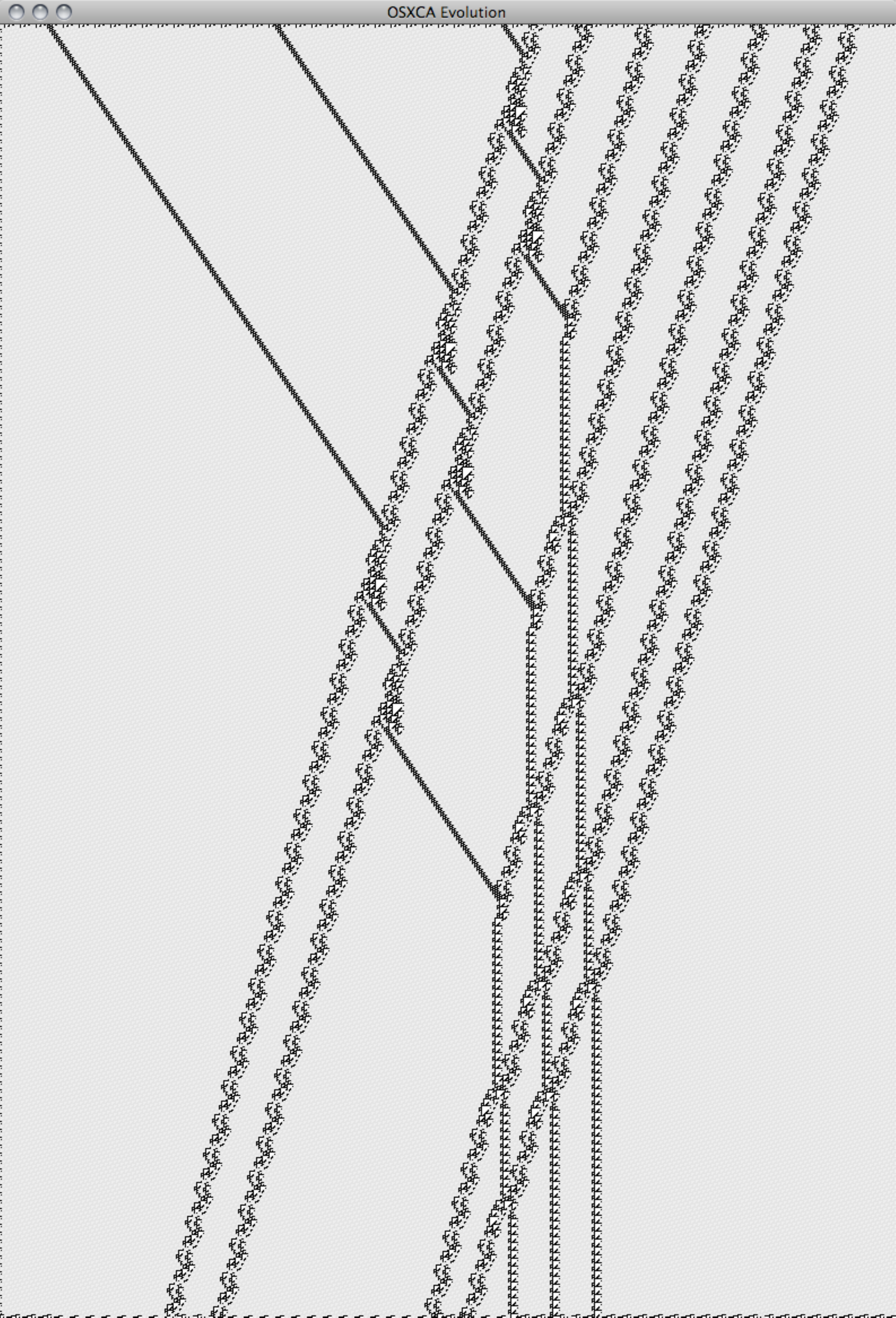}}
\caption{Solitons working to yield, handle, and control bits in one cyclic tag system working in Rule 110. This evolution begins with 793 cells to 1144 generations, the evolution is filtered suppressing its ether (periodic background).}
\label{cookObject}
\end{figure}

Figure~\ref{cookObject} summarizes the function of solitons in the construction of a cyclic tag system in Rule 110. The first important fact is the solitonic reaction between packages of $A^{4}$ mobile self-localizations versus one $\bar{E}$ mobile self-localization, this collision should introduce bits on the tape of the cyclic tag system yielding a series of $C_2$ mobile self-localizations, also it might also avoid destroying extra $\bar{E}$ mobile self-localizations that cannot infer on the computable system. Besides, when a binary digit is into the tape (stationary $C_2$ mobile self-localizations) hence a package of $\bar{E}$ mobile self-localizations, coming from the right side shall preserve the binary digit and across them with other solitonic collision, because forward them self will be transformed in binary data. Here we can see how solitons are useful to preserving information and recognising when a value will be processed, deleted, or read (for full details please see \cite{kn:MMS11}).

\section{Conclusions}
\label{conclusions}

We have reported a complete number of solitons in ECA Rule 110 from binary collisions. For ECA Rule 54 we have reported a new number of collisions that yield solitons, where they could be manipulated to develop computable devices, or yet more, complex constructions based solitonic reactions. So, finally we have characterized a simple single-soliton two-component solution with a simple ECAM $\phi_{R9maj:4}$, where mobile self-localizations always work as solitons even starting from random initial conditions, because each soliton is always constructed from $\phi_{R9maj:4}$.

With regards to memory effect in CA \cite{kn:Alo09, kn:Alo11}. Recently, we have studied how a memory function helps to describe dynamics properties that are not evident at the first instance \cite{kn:MAA10, kn:MAA, kn:MAS10, kn:MAA12}. In present paper, the majority memory selected in ECA Rule 9 opens a new evolution rule ECAM $\phi_{R9maj:4}$ able to simulate solitons. Of course, fragments of the original evolution rule determine such dynamics.


\end{document}